\documentclass[12pt]{article}

\def\bphi{\mbox{\boldmath $\phi$}}

\def\balpha{\mbox{\boldmath $\alpha$}}

\def\btau{\mbox{\boldmath $\tau$}}
\def\bnu{\mbox{\boldmath $\nu$}}
\def\bpsi{\mbox{\boldmath $\psi$}}
\def\brho{\mbox{\boldmath $\rho$}}
\def\bgamma{\mbox{\boldmath $\gamma$}}
\def\bchi{\mbox{\boldmath $\chi$}}
\def\bm{\mbox{\bf m}}

\def\phivac{\Phi_{\mathrm{\scriptscriptstyle{VAC}}}}

\input{amssymb.sty}

\def\bbr{{\mathbb R}}

\def\tr{\mathop{\rm tr}\nolimits}

\def\frac#1#2{{{#1}\over{#2}}}
\def\tfrac#1#2{{\textstyle{{#1}\over{#2}}}}

\def\half{\tfrac{1}{2}}
\def\third{\tfrac{1}{3}}

\begin{document}

\begin{titlepage}

\baselineskip 24pt

\begin{center}

{\Large {\bf Developing the Framed Standard Model}}

\vspace{.5cm}

\baselineskip 14pt

  {\large Michael J BAKER and Jos\'e BORDES}\\
michael.baker@uv.es and jose.m.bordes\,@\,uv.es\\
{\it Departament Fisica Teorica and IFIC, Centro Mixto CSIC, Universitat de Valencia,
  Calle Dr. Moliner 50, E-46100 Burjassot (Valencia), Spain\footnote{Work supported by Spanish MICINN under contracts Proyecto Prometeo 2008-004 CI11-086 and FPA2008-02878.}}\\
\vspace{.2cm}
{\large CHAN Hong-Mo}\\
h.m.chan\,@\,stfc.ac.uk \\
{\it Rutherford Appleton Laboratory,\\
  Chilton, Didcot, Oxon, OX11 0QX, United Kingdom}\\
\vspace{.2cm}
{\large TSOU Sheung Tsun}\\
tsou\,@\,maths.ox.ac.uk\\
{\it Mathematical Institute, University of Oxford,\\
  24-29 St. Giles', Oxford, OX1 3LB, United Kingdom}

\end{center}

\vspace{.3cm}

\begin{abstract}

The framed standard model (FSM) suggested earlier, which
incorporates the Higgs field and 3 fermion generations 
as part of the framed gauge theory structure, is here 
developed further to show that it gives both quarks
and leptons hierarchical masses and mixing matrices akin 
to what is experimentally observed.  Among its many 
distinguishing features which lead to the above results 
are (i) the vacuum is degenerate under a global $su(3)$ 
symmetry which plays the role of fermion generations, 
(ii) the fermion mass matrix is ``universal'', rank-one 
and rotates (changes its orientation in generation space) 
with changing scale $\mu$, (iii) the metric in generation 
space is scale-dependent too, and in general non-flat, 
(iv) the theta-angle term in the QCD action of topological 
origin gets transformed into the CP-violating phase of the 
CKM matrix for quarks, thus offering at the same time a 
solution to the strong CP problem. 
 
\end{abstract}

\end{titlepage}

\clearpage

\section{Introduction}

Despite its great success in explaining existing data, the 
standard model as usually formulated is based on a number of
intricate assumptions some of which are themselves in need of
explanation.  These include in particular the assumption of
the scalar Higgs field needed for symmetry breaking in the 
electroweak sector and the introduction of 3 generations of
fermion fields, neither of which has a theoretical foundation
in a theory otherwise quite geometrically grounded.  Even more
mysterious is the injection from experiment of the hierarchical
fermion mass spectrum and the peculiar mixing pattern between
up and down fermion states, which together account for some 
two-thirds of the model's twenty-odd empirical parameters.  For 
these reasons, among others, it is generally expected that a more fundamental 
theory exists from which the present standard model can be 
derived with all the mysterious patterns that it contains.

What we now call the framed standard model (FSM) is an attempt
initiated in \cite{prepsm} at constructing just such a theory.  
It is based on what one can call the framed gauge theory (FGT) 
framework \cite{efgt} in which, in addition to the gauge and 
matter fields in standard formulations of gauge theories, one 
also includes the frame vectors in internal symmetry space 
as dynamical variables.  That frame vectors can appear as 
dynamical variables is familiar in gravity, where 
vierbeins are often taken as such in place of the metric.  So 
it seems reasonable to consider this possibility also in the 
particle physics context.  The immediate attraction of this is 
three-fold.  First, the frame vectors in internal symmetry space 
transform as fundamental representations of the gauge symmetry 
but are Lorentz scalars, and so can function as the Higgs fields 
needed for breaking the flavour symmetry.  Secondly, they carry 
by definition, in addition to indices referring to the local 
gauge frame, ``dual'' indices referring to a global reference 
frame which, in the case of colour $su(3)$ symmetry, can 
play the role of fermion generations.  Thirdly, since physics 
should be independent of the choice of the reference frame, the 
action containing these frame vectors (or ``framons'') as dynamical 
variables should be invariant under both the global ``dual'' symmetry and the original local gauge symmetry, thus greatly 
reducing the freedom in the form that the action can take.  Indeed, 
it was shown in \cite{prepsm}, and more succinctly and
with greater transparency in \cite{efgt}, that applying this idea to 
a theory with the gauge symmetry $su(3)\times su(2) \times u(1)$ 
yields a structure, namely the framed standard model (FSM), 
which is standard model-like, but now with both the Higgs field 
and 3 fermion generations already built in, as desired.
 
The purpose of the present paper is to develop further this FSM 
to see whether it could yield for the 3 generations of fermions 
a hierarchical mass spectrum and mixing matrices both with the 
distinctive features experimentally observed.  Our strategy for 
doing so is based on the observation that if the fermion mass 
matrix is of rank one, ``universal'', and rotates (i.e., changes 
its orientation in generation space) with scale, then both mass 
hierarchy and mixing will automatically result \cite{r2m2}.  It 
has already been shown in \cite{prepsm,efgt} that in the FSM, 
the fermion mass matrix is indeed of rank one, being expressible 
in a factorized form:
\begin{equation}
m = m_T {\balpha}{\balpha}^{\dagger},
\label{mfact}
\end{equation}
in terms of a unit vector $\balpha$ in generation space, and
is ``universal'' in the sense that the vector $\balpha$ is the 
same for all fermion types $T$, i.e., whether up or down, and 
whether leptons or quarks.  Rotation of the mass matrix then 
just means that this $\balpha$ rotates.  In the next section \S2,
we shall outline for convenient reference how rotation works,
and summarize some of the formulae needed for the discussion later.
The main thrust of the paper, however, is directed towards the
question whether the FSM will generate the rotation we seek, and
if it does how it manages to do so.

Notice that neither the ideas of a rank-one mass matrix nor 
of its rotation with scale are new.  The experimentally observed 
facts that fermion masses are hierarchical and that they mix with 
a mixing matrix close to unity (at least for quarks) has 
long suggested to some authors \cite{Fritsch,Harari} that a 
rank-one mass matrix would be phenomenologically profitable.  It
has also long been known \cite{Ramond} that even in the standard 
model as usually formulated the fermion mass matrix rotates
with changing scales as a consequence of up-down mixing.  Hence,
the only really new  concept here is that instead of mixing giving
rise to rotation as in \cite{Ramond}, it is rotation that gives
rise to mixing, requiring thus a faster rotation than is given in
the standard model by mixing.    
That one has to go beyond the standard model for the
rotation we seek for explaining mass hierarchy and mixing is obvious since,
in the standard model itself, masses and mixing angles appear as empirical
parameters, meaning that the standard model is internally consistent for any
choice of their values, and therefore inherently incapable of explaining
them.

The manner that this faster rotation comes about in the FSM is 
in fact quite intriguing and deserves a brief outline before we
plunge into details.  As shown in \cite{prepsm,efgt}, to which 
the reader is referred for details, the FSM can be thought of as
consisting of two sectors, the electroweak and the strong, each 
with its own set of gauge vector bosons and frame-vector scalar 
bosons (framons).  What is new compared to the standard model are 
of course the framons and in particular their self-interaction
term $V[\Phi]$, which will determine, by its minimization, the 
vacuum.  The form that $V[\Phi]$ can take is severely constrained 
by invariance principles, as we already stated; its explicit form 
derived in \cite{prepsm,efgt} will be given below.  For the moment, 
we need only note that $V[\Phi]$ consists of 3 terms, a term $V_{W}$ 
depending only on the weak framon field which is essentially the 
same as for the standard electroweak theory on which little need 
at present to be said, then a term $V_S$ depending only on the 
strong framon field, and lastly a linkage term $V_{WS}$ depending 
on both the weak and strong framon fields.  From these, one can 
deduce the following.  First, $V_S$ alone on minimization gives a 
vacuum where the 3 strong framons form an orthonormal triad, as 
frame vectors are expected to do, but they are distorted from 
orthonormality by the linkage term $V_{WS}$, with the distortion 
depending on the direction of a vector ${\balpha}$ coming from 
the weak sector, and this vector is exactly that which appears in 
the rank-one fermion mass matrix expressible as in (\ref{mfact}).
Secondly, when loop corrections are turned on, the vacuum will get 
renormalized and will change with scale carrying the vector ${\balpha}$ 
along with it.  And it is this change in direction of ${\balpha}$ 
(rotation) which gives in the end the mass hierarchy and mixing 
patterns one wants.  One sees that the rotation here actually 
originates in the vacuum and gets transmitted to the fermion 
mass matrix only through the vector $\balpha$.  It is therefore 
independent of the fermion type $T$ to which $\balpha$ is coupled, 
meaning that the mass matrix will remain of the factorized form 
(\ref{mfact}) above and universal under rotation, both 
conditions needed for the rotation scheme \cite{r2m2} to work.
Moreover, one sees that it is the strong interactions which is 
driving the rotation, which can therefore be fast enough to give
the mixing effects one seeks.

To examine in detail the process outlined above, one will need to 
go successively through the following steps.  First, of course, 
one will need
\begin{itemize}
\item \S3 to elucidate the vacuum as derived from minimizing 
$V[\Phi]$.
\end{itemize}
Next, one will need to examine how this vacuum gets renormalized,
and hence becomes scale-dependent.  In principle, one can obtain
information on this via the renormalization on any quantity which 
depends on the vacuum.  We have chosen in this paper, mainly for 
historical reasons, to study the mass renormalization of certain
fermion states to be specified later which are, in the present 
FSM framework, the analogues in the strong sector of leptons and 
quarks in the electroweak sector.  They are hadron states.  And as 
a sample of the renormalization effects on these, we have chosen 
to study those due to the insertion of a loop of what we shall 
call strong Higgs states which were found in a parallel earlier
study \cite{ckm} to give most of the rotation.  These strong Higgs 
states are fluctuations of the strong framons about the strong 
vacuum, and are thus analogues in the strong sector of the standard 
Higgs state in the electroweak sector.  To achieve this aim, one 
will need
\begin{itemize} 
\item \S4 to identify the strong Higgs states from framon fluctuations 
about the vacuum;
\item \S5 to derive their couplings to the chosen fermion states 
and evaluate their loop correction to the fermion self-energy, and 
hence to derive its implication on the scale-dependence of the 
vacuum;
\item \S6 to derive the rotation equation resulting from this for the 
vector ${\balpha}$ appearing in (\ref{mfact}) above. 
\end{itemize}
Some of these steps were started in \cite{prepsm} but not completed
or were done only to a rough approximation.  Now, with much better
techniques and improved understanding, the programme set out above can 
be carried out exactly and in full.  

The resulting rotation equation for $\balpha$, though likely to be 
rather limited by its mode of derivation in accuracy and range of 
applicability, serves nevertheless as a useful concrete example 
for how rotation in the FSM is generated.  As already noted, the 
rotation still leaves the fermion mass matrix both factorized and 
universal as required.  Besides, it is seen
\begin{itemize}
\item \S7 to have fixed points at $\mu = 0, \infty$;
\item \S9 to generate automatically a CP-violating phase in the CKM
matrix and to offer, at the same time, a solution to the strong CP 
problem;
\end{itemize}
properties which are believed to be generic, i.e., independent of 
much of the restrictive assumptions under which the rotation is 
derived here.  In other words, the FSM  seems to possess already 
those features which have been identified in the phenomenological 
study \cite{r2m2} as needed of a rotational model for a successful 
description of the mass and mixing data.  

One other novel feature of the FSM revealed in the analysis of the
rotation equation is the appearance of
\begin{itemize}  
\item \S8 a running metric for generation space,
\end{itemize}
which is of much theoretical, and perhaps even phenomenological,
interest for the future, but is shown not to alter the other 
effects of rotation already listed.

A brief summary of the results and comparisons to other models 
beyond the standard model is given in the last section \S10.

\section{Mass Hierarchy and Mixing from Rotation}

We begin, for easy reference and to introduce some notations, 
with a brief outline of how a rank-one rotating mass matrix 
(R2M2) automatically leads to mass hierarchy and to mixing 
between up and down fermion states, while displaying several 
formulae useful for later discussions.  For details, the reader 
is referred to \cite{r2m2}, a recent review.

We note first that any fermion mass matrix can, by a judicious 
relabelling of the $su(2)$ singlet right-handed fields, be cast 
into a form with no dependence on $\gamma_5$ \cite{Weinbergren} so 
that any rank-one mass matrix can be written without loss of 
generality in the form (\ref{mfact}).  Then the assertion that 
$m$ rotates simplifies to the assertion
that the vector ${\balpha}$ rotates.  

That an ${\balpha}$ rotating with scale will automatically lead 
to mixing and mass hierarchy is most easily explained in the 
simplified situation when account is taken only of the two 
heaviest generations.  By (\ref{mfact}) then, taking for the 
moment ${\balpha}$ to be real and $m_T$ to be $\mu$-independent 
for simplicity, we would have $m_t = m_U$ as the mass of $t$ 
and the eigenvector ${\balpha}(\mu = m_t)$ as its state vector 
${\bf t}$.  Similarly, we have $m_b = m_D$ as the mass and 
${\balpha}(\mu = m_b)$ as the state vector ${\bf b}$ of $b$. 
The vectors ${\bf t}$ and ${\bf b}$ are not aligned, being 
the vector ${\balpha}(\mu)$ taken at two different values of 
its argument $\mu$, and ${\balpha}$ by assumption rotates.  Let 
then $ \theta_{tb}$ be the non-zero angle between them.  Next, 
the state vector ${\bf c}$ of $c$ must be orthogonal to ${\bf t}$, 
$c$ being by definition an independent quantum state to $t$.  
Similarly, the state vector ${\bf s}$ of $s$ is orthogonal to 
${\bf b}$.  The up dyad $\{{\bf t}, {\bf c}\}$ differs thus from the
down dyad $\{{\bf b}, {\bf s}\}$ by a rotation by the angle 
$\theta_{tb}$ above.  This gives then the following CKM mixing 
(sub)matrix in the situation with only the two heaviest states
being considered
\begin{equation}
\left( \begin{array}{cc} V_{cs} & V_{cb} \\ V_{ts} & V_{tb} \end{array}
   \right) = \left( \begin{array}{cc}  {\bf c} \cdot{\bf s}  
                             &  {\bf c} \cdot{\bf b}  \\
                                {\bf t} \cdot{\bf s} 
                             &  {\bf t} \cdot{\bf b}
             \end{array} \right )
           = \left( \begin{array}{cc} \cos \theta_{tb} & -\sin \theta_{tb} \\  
                \sin \theta_{tb} & \cos \theta_{tb} \end{array} \right),    
\label{UDmix2}
\end{equation}
which is no longer the identity, hence mixing.    

Next, what about hierarchical masses?  From (\ref{mfact}), it follows 
that ${\bf c}$ must have zero eigenvalue at $\mu = m_t$.  But this 
value is not to be taken as the mass of $c$ which has to be measured 
at $\mu = m_c$.  In other words, $m_c$ is to be taken as the 
solution to the equation
\begin{equation}
\mu = \langle {\bf c}|m(\mu)|{\bf c} \rangle 
    = m_U |\langle {\bf c}|{\balpha}(\mu) \rangle|^2.
\label{solvmc}
\end{equation}
A non-zero solution exists since the scale on the LHS decreases from 
$\mu=m_t$ while the RHS increases from zero at that scale.  Another
way to see this is that since ${\balpha}$ by assumption rotates
so that at $\mu < m_t$, it would have rotated to some direction
different from ${\bf t}$, and acquired a component, say 
$\sin \theta _{tc}$, in the direction of ${\bf c}$ giving thus
\begin{equation}
m_c = m_t \sin^2 \theta_{tc},
\label{mc2}
\end{equation}
which is non-zero but will be small if the rotation is not too fast,
hence mass hierarchy.  

That the mass spectra and mixing matrices so obtained from 
a rank-one rotating mass matrix (R2M2) actually do resemble 
those observed in experiment is also readily checked in the
present simplification when only the two heaviest states of
each quark type are considered.  By inverting the above simple 
formulae (\ref{mc2}) and (\ref{UDmix2}) for the masses and 
mixing angles, one easily derives the corresponding values of
the angle $\theta$ at various scales.  If R2M2 is indeed
valid, then these values should all fall on a smooth curve as
a function of $\mu$ representing the rotation trajectory for the
vector $\balpha$.  This exercise performed in \cite{cevidsm} gave
results very well fitted by an exponential which showed that the then 
available data were fully consistent with the hypothesis.

One sees then that with 2 generations, rotation will give
automatically both mixing and mass hierarchy, as claimed.
Basically the same argument is applicable to the realistic
3-generation case, though the analysis becomes a little 
more intricate.  We give here only the result for future
reference, the detailed derivation of which can be found
in, for example, \cite{r2m2}.  For $U$-type quarks, the
state vectors are defined in terms of the rotating vector
${\balpha}$ via
\begin{eqnarray}
{\bf t} & = & {\balpha}(m_t), \nonumber \\
{\bf c} & = & {\bf u} \times {\bf t}, \nonumber \\
{\bf u} & = & \frac{{\balpha}(m_t) \times {\balpha}(m_c)}
   {|{\balpha}(m_t) \times {\balpha}(m_c)|}.
\label{Utriad}
\end{eqnarray}
And their masses are given by
\begin{eqnarray}
m_t & = & m_U, \nonumber \\
m_c & = & m_U |{\balpha}(m_c) \cdot{\bf c}|^2, \nonumber \\
m_u & = & m_U |{\balpha}(m_u) \cdot{\bf u}|^2.
\label{hiermass}
\end{eqnarray}

Together, these 2 sets of coupled equations allow us to
evaluate both the state vectors and the masses.  Similar
equations and remarks apply also to $D$-type quarks.  With the state
vectors so determined, the mixing matrices could then be 
directly evaluated, e.g., for quarks \cite{Cabibbo,KM}
\begin{equation}
V_{\rm CKM} \sim \left( \begin{array}{ccc}
   {\bf u} \cdot{\bf d}  &  {\bf u} \cdot
{\bf s}  &  {\bf u} \cdot{\bf b}  \\
    {\bf c} \cdot{\bf d}  &  {\bf c} \cdot
{\bf s}  &  {\bf c} \cdot{\bf b}  \\
    {\bf t} \cdot{\bf d}  &  {\bf t} \cdot
{\bf s}  &  {\bf t} \cdot{\bf b}  
          \end{array} \right).
\label{VCKM}
\end{equation}
The expression for the lepton mixing matrix $U_{\rm PMNS}$ \cite{ponte,mns}
would be similar. 

That the mass spectra and mixing matrices so obtained from 
R2M2 are still consistent with experiment when all 3 
generations of fermions are taken into account is shown in 
\cite{btfit} and is reviewed 
in \cite{r2m2} to which the reader is referred.

In the above summary, it has been assumed that all the vectors
in generation space are real and that their norms and products
are calculated with a flat metric. It will be seen later that 
if account is taken of the theta-angle term in the QCD action,
famous in the old strong CP problem \cite{Weinbergbook}, 
then the vectors can
become complex, giving $V_{\rm CKM}$ a CP-violating phase \cite{KM}.  Also, 
as the present model (FSM) develops, it will be seen that the 
metric in generation space can become distorted from flatness.  
However, it will be shown that even in these circumstances, 
all the formulae listed above, only with certain provisos, 
will still remain valid.

\section{The Vacuum}

We turn now to our main task of seeing how rotation develops in
the FSM, beginning with an elucidation of the vacuum. 

The formulation in \cite{prepsm,efgt} of the FSM as the ``minimally 
framed'' gauge theory, i.e., the framed gauge theory with the 
smallest number of scalar framon fields, for the gauge symmetry 
$su(3) \times su(2) \times u(1)$, gives two types of framons: a 
``weak framon'' field of the form
\begin{equation}
\phi_r^{\tilde{r} \tilde{a}} = \alpha^{\tilde{a}} \phi_r^{\tilde{r}},
    \ \ r, \tilde{r} = 1, 2, \ \  \tilde{a} = 1, 2, 3, \ \ \ \ y = 
\pm \half, \ \
    \tilde{y} = \mp \half,
\label{wframon}
\end{equation}
and a ``strong framon'' field of the form
\begin{equation}
\phi_a^{\tilde{r} \tilde{a}} = \beta^{\tilde{r}} \phi_a^{\tilde{a}},
    \ \ \tilde{r} = 1, 2, \ \ a, 
\tilde{a} = 1, 2, 3, \ \ \ \  y = - \third, \ \
    \tilde{y} = \third.
\label{sframon}
\end{equation}
where $\phi_r^{\tilde{r}}$ and $\phi_a^{\tilde{a}}$ are scalar 
space-time $x$-dependent fields, while the factors ${\alpha}
^{\tilde{a}}$ and ${\beta}^{\tilde{r}}$are global $x$-independent 
quantities and $y$ and $\tilde{y}$ denote the $u(1)$ and 
$\tilde{u}(1)$ charges respectively.  The components
$\phi_r^{\tilde{r}}$ and $\phi_a^{\tilde{a}}$ are not all
independent; the weak framons satisfy
\begin{equation}
\phi_r^{\tilde{2}} = - \epsilon_{rs} (\phi_s^{\tilde{1}})^*,
\label{su2ortho}
\end{equation}
while the strong framons satisfy
\begin{equation}
{\rm det} (\Phi) = ({\rm det} (\Phi))^*,
\label{detreal}
\end{equation}
where we have arranged the strong 
framon fields $\phi_a^{\tilde{a}}$ as a matrix, $\Phi = 
(\phi_a ^{\tilde{a}})$.  Furthermore, we introduce the 2-vector
$\bphi$ as a shorthand notation for the single weak framon 
$\phi^{\tilde{1}}_r$ (see equation (\ref{su2ortho})). 

Since physics should be independent of the choice either of the 
local or the global reference frame, the action constructed with 
these framon fields carrying both local and global indices has 
to be invariant under both the original local gauge symmetry 
$su(3) \times su(2) \times u(1)$ and its ``dual'', the global symmetry 
$\widetilde{su}(3) \times \widetilde{su}(2) \times \tilde{u}(1)$, 
which places severe restrictions on the form it can 
take.  In particular, the self-interaction term of the framons, which
we call the framon potential $V[\Phi]$, is restricted by invariance 
plus renormalizability to the form \cite{prepsm,efgt}
\begin{equation}
V[\Phi] = V_W[\Phi] + V_S[\Phi] + V_{WS}[\Phi]\,,
\label{VPhi}
\end{equation}
where $V_W$ involves only the weak framons, $V_S$ only the strong 
framons, and $V_{WS}$ both, with
\begin{equation}
V_W[\Phi] = - \mu_W |\bphi|^2 + \lambda_W (|\bphi|^2)^2,
\label{VPhiW}
\end{equation} 
\begin{equation}
V_S[\Phi] = - \mu_S \sum_{a,\tilde{a}} (\phi_a^{\tilde{a}*}\phi_a^{\tilde{a}})
    + \lambda_S \left[ \sum_{a, \tilde{a}} (\phi_a^{\tilde{a}*}
    \phi_a^{\tilde{a}}) \right]^2 + \kappa_S \sum_{a,b,\tilde{a},\tilde{b}}
    (\phi_a^{\tilde{a}*} \phi_a^{\tilde{b}})
    (\phi_b^{\tilde{b}*} \phi_b^{\tilde{a}}),
\label{VPhiS}
\end{equation}
and
\begin{equation}
V_{WS}[\Phi] = \nu_1 |\bphi|^2 \sum_{a,\tilde{a}} \phi_a^{\tilde{a} *}
    \phi_a^{\tilde{a}} - \nu_2 |\bphi|^2 \sum_a \left| \sum_{\tilde{a}}
    (\alpha^{\tilde{a} *} \phi_a^{\tilde{a}})\, \right|^2.
\label{VPhiWS}
\end{equation}
The potential $V[\Phi]$ depends on 7 real coupling parameters in
all, namely $\mu_W, \lambda_W, \mu_S, \lambda_S, \kappa_S, \nu_1,
\nu_2$.  Although these parameters can in principle have either
sign, we take $\mu_W, \lambda_W, \mu_S, \lambda_S$ all to be 
positive so that both the weak and strong vacua are degenerate,
and also $\kappa_S$ to be positive for reasons which will soon 
be apparent.  The other 2 parameters $\nu_1, \nu_2$, however, 
can have either sign in the following discussion.  

The object of this section is to identify the framon vacuum by 
minimizing this framon potential.  This can be done, of course, 
by fixing first a gauge each for both the local and the global 
symmetry, then differentiating the potential with respect to the 
remaining 12 variables and putting the derivatives to zero.  It
will be straightforward, but rather complicated.  
The reason is that although the 
strong potential $V_S[\Phi]$ by itself gives a minimum for which 
the strong framons remain orthonormal, the term $V_{WS}[\Phi]$ 
which links the strong and weak sectors distorts the vacuum 
values of the strong framons from orthonormality, and it is
the necessity of referring to these non-orthonormal frames which
makes the analysis complicated.  This will especially be the 
case when we are interested in tracing the scale-dependence 
of the vacuum, i.e., how the vacuum moves from one to another 
among the degenerate set, when all these vacua are distorted 
from orthonormality each in a different way.

For this reason, we adopt here a different tack.  A point to note
first is that the degeneracy of the strong vacuum originates from 
the invariance of the potential $V_S$ under the global symmetry 
$\widetilde{su}(3)$.  When we arrange the strong framon fields 
$\phi_a^{\tilde{a}}$ above as a matrix $\Phi$, 
with $a$ labelling the rows and $\tilde{a}$
the columns, then the $\widetilde{su}(3)$ transformations are 
represented by unitary matrices, say $A^{-1}$, operating from 
the right.  Now if $\Phi$ at vacuum is orthonormal, then so 
is $\phivac A^{-1}$, but if $\phivac$ is not orthonormal, 
$\phivac A^{-1}$ can take many different shapes.  Nevertheless, 
any two of all these differently shaped vacua are still related 
just by some $A^{-1}$ from $\widetilde{su}(3)$.\footnote{A point 
already noted in \cite{prepsm}, though not then fully utilized.}

With this realization one sees that one need not actually perform 
the minimization analysis around a general vacuum, but only around 
a vacuum for which the analysis is particularly simple, since any 
other vacuum can be obtained from it by applying the appropriate 
$\widetilde{su}(3)$ transformation.   Equivalently, we can think
of choosing for any given vacuum an appropriate gauge so as to 
make it appear particularly simple, and all other vacua in the same
gauge can be obtained from it by $\widetilde{su}(3)$ transformations. 
Indeed, one will find that the analysis becomes then so simple that 
any vacuum can be found in this way even without doing any actual 
minimization.

Suppose then we choose or are given some particular vacuum in the 
degenerate set as our reference vacuum, quantities defined with
respect to which we shall indicate by an index (either as subscript 
or superscript) 0.  It will correspond to some value of the vector 
${\balpha}$, say ${\balpha}_0$.  Let us choose to work in the 
$\widetilde{su}(3)$ gauge where the vector ${\balpha}_0$ is real 
and points in the first direction.  
This does not fix the $\widetilde{su}(3)$ gauge completely, but we 
can leave the rest unspecified for the moment.  We ask now how the 
chosen reference vacuum will appear in this new $\widetilde{su}(3)$
gauge.  To be specific, we shall need also to fix a gauge for the
local $su(3)$ symmetry.  We can choose \cite{prepsm},\ e.g., either the
triangular gauge (where the framon matrix elements are real along the 
diagonal and vanishing below it) or the hermitian gauge (where 
the framon matrix  $\Phi$ is hermitian), but with ${\balpha_0}
= (1,0,0)$ the two gauges coincide.

To find the values of the framons $\Phi$ at the reference vacuum 
in the chosen gauge, let us rewrite the potential $V[\Phi]$ in 
(\ref{VPhi}) in terms of the vectors ${\bphi}^{\tilde{a}}=(\phi^{\tilde{a}}_a)$ 
in $su(3)$ space.  For $\balpha=(1,0,0)$, the potential then takes the
form
\begin{eqnarray}
V[\Phi] & = & - \mu_W |\bphi|^2 + \lambda_W (|\bphi|^2)^2
          - \mu_S \sum_{\tilde{a}} |{\bphi}^{\tilde{a}}|^2
          + \lambda_S \left( \sum_{\tilde{a}}
            |{\bphi}^{\tilde{a}}|^2 \right)^2 \nonumber \\
& & + \kappa_S \sum_{\tilde{a}} \left( |{\bphi}^{\tilde{a}}|^2
                                              \right)^2  
    + \kappa_S \sum_{\tilde{a} \neq \tilde{b}}
    |{\bphi}^{\tilde{a}*}\cdot{\bphi}^{\tilde{b}}|^2 \nonumber \\
& &   + \nu_1 |\bphi|^2 \sum_{\tilde{a}} |{\bphi}^{\tilde{a}}|^2
    - \nu_2 |\bphi|^2 |{\bphi}^{\tilde{1}}|^2.
\label{VPhia}
\end{eqnarray}
We note in this that only the second $\kappa_S$ term depends on
the orientation of the vectors ${\bphi}^{\tilde{a}}$, the other
terms depending only on their lengths.  Hence, minimizing $V[\Phi]$
with respect to their orientations, we obtain (for $\kappa_S > 0$) 
that these vectors will be mutually orthogonal at minimum, so that 
this second $\kappa$ term will be zero and can be dropped from 
consideration.

Next, let us rewrite the $\nu_2$ term as
\begin{equation}
\nu_2 |\bphi|^2 \left[ \left( -\frac{2}{3} |{\bphi}^{\tilde{1}}|^2
                             +\frac{1}{3} |{\bphi}^{\tilde{2}}|^2
                             +\frac{1}{3} |{\bphi}^{\tilde{3}}|^2
                             \right)
  -\frac{1}{3} \sum_{\tilde{a}} |{\bphi}^{\tilde{a}}|^2 \right],
\label{nu2term}
\end{equation}
where the second term has the same form as, and can be absorbed into, 
the $\nu_1$ term in $V[\Phi]$.  Combining now the remaining terms in 
(\ref{nu2term}) with the remaining $\kappa_S$ term in $V[\Phi]$ by 
completing squares, we can rewrite the sum as
\begin{equation}
\kappa_S \left[ \left( \sum_{\tilde{a}} |{\bphi}'^{\tilde{a}}|^2
  \right)^2 - \frac{1}{6} \frac{\nu_2^2}{\kappa_S^2} (|\bphi|^2)^2
  \right],
\label{newkappaterm}
\end{equation}
with
\begin{equation}
|{{\bphi}'}^{\tilde{1}}|^2 = |{\bphi}^{\tilde{1}}|^2
               -\frac{1}{3} \frac{\nu_2}{\kappa_S} |\bphi|^2, \ \ \ 
|{{\bphi}'}^{\tilde{2}}|^2 = |{\bphi}^{\tilde{2}}|^2
               +\frac{1}{6} \frac{\nu_2}{\kappa_S} |\bphi|^2, \ \ \ 
|{{\bphi}'}^{\tilde{3}}|^2 = |{\bphi}^{\tilde{3}}|^2
               +\frac{1}{6} \frac{\nu_2}{\kappa_S} |\bphi|^2.
\label{phiprime}
\end{equation}
Again the last term in (\ref{newkappaterm}) has the same form as,
and can be absorbed into, the $\lambda_W$ term in $V[\Phi]$.
 
Noting that
\begin{equation}
\zeta_S^2 = \sum_{\tilde{a}} |{{\bphi}'}^{\tilde{a}}|^2
   = \sum_{\tilde{a}} |{{\bphi}}^{\tilde{a}}|^2  
\label{sumphieq}
\end{equation}
we see that the potential as a whole now resembles the old potential
without the $\nu_2$ term, only with $|{{\bphi}}^{\tilde{a}}|^2$
replaced by $|{{\bphi}'}^{\tilde{a}}|^2$ and some changes in the
definition of $\nu_1$ and $\lambda_W$.  In particular, we note that
the potential is symmetric in $|{{\bphi}'}^{\tilde{a}}|^2$ so
that even without differentiation we can conclude that the minimum
is at
\begin{equation}
|{\bphi}'^{\tilde{1}}|^2 = |{\bphi}'^{\tilde{2}}|^2
   = |{\bphi}'^{\tilde{3}}|^2 = \frac{\zeta_S^2}{3},
\label{vacphip}
\end{equation}
or at
\begin{equation}
|{\bphi}^{\tilde{1}}|^2 = \zeta_S^2 (\tfrac{1+2R}{3}), \ \ \ 
|{\bphi}^{\tilde{2}}|^2 = \zeta_S^2 (\tfrac{1-R}{3}), \ \ \ 
|{\bphi}^{\tilde{3}}|^2 = \zeta_S^2 (\tfrac{1-R}{3}),
\label{vacphi}
\end{equation}
with
\begin{equation}
R = \frac{\nu_2 \zeta_W^2}{2 \kappa_S \zeta_S^2}.
\label{R}
\end{equation}
It follows then immediately that at ${\balpha} = (1,0,0)$, 
 $\phivac^0$ is necessarily diagonal because of the mutual 
orthogonality of the vectors ${\bphi}^{\tilde{a}}$, and that it 
will take the simple form
\begin{equation}
\phivac^0 = V_0^0 = \zeta_S \left( \begin{array}{ccc} 
                   \sqrt{\frac{1+2R}{3}} & 0 & 0 \\
                   0 & \sqrt{\frac{1-R}{3}} & 0 \\
                   0 & 0 & \sqrt{\frac{1-R}{3}} \end{array} \right).
\label{Phivac0}
\end{equation}

That the vacuum at ${\balpha} = (1,0,0)$ should take this form 
is actually, {\it a posteriori}, fairly obvious.  We recall that
the strong potential $V_S$ would by itself imply vacuum values
for $\bphi^{\tilde{a}}$ which are mutually orthogonal and of
equal lengths, and it is the term $V_{WS}$ which distorts them
from orthonormality.  Choosing then $\balpha$ to point in the
$(1,0,0)$ direction means that only the lengths of the vectors
will be affected, and this effect will depend on the relative
strengths of the $\kappa_S$ and $\nu_2$ terms, namely on the
parameter $R$ in (\ref{R}) above.

Having now (\ref{Phivac0}) for the reference vacuum, one can 
obtain $\phivac$ for any other vacuum in the degenerate set, 
say one corresponding to a different vector  $\balpha$, by 
an $\widetilde{su}(3)$ transformation $A^{-1}$, applied from 
the right, thus
\begin{equation}
\phivac = V_0 = \phivac^0\, A^{-1},
\label{Phivac}
\end{equation}
where
\begin{equation}
{\balpha} = A \left( \begin{array}{c} 1 \\ 0 \\ 0 \end{array}
                 \right),
\label{alpha}
\end{equation}
though still in the same gauges as before.  We notice that 
(\ref{Phivac}) is in general not diagonal, meaning that the 
framon vectors $\bphi^{\tilde{a}}$ at a general vacuum are now 
neither mutually orthogonal nor similarly normalized in the 
chosen gauges.  Indeed, $\phivac$ will in general not even 
be triangular nor hermitian, but can be made to be either by 
an appropriate change in the local gauge, i.e., by operating 
with an $su(3)$ transformation from the left \cite{prepsm}. 
Further, of course, $\phivac$ can be transformed back into 
the canonical form (\ref{Phivac0}) by appropriate gauge changes
in both local $su(3)$ and global $\widetilde{su}(3)$.  But in
what follows, unless otherwise stated, we shall keep working 
in the chosen gauges as before where $\phivac^0$ is
of the form (\ref{Phivac0}).  

From (\ref{Phivac}), it follows that
\begin{equation}
\phivac^\dagger \phivac 
   = \zeta_S^2 (\tfrac{1 + 2R}{3}) P_{\balpha} 
        + \zeta_S^2 (\tfrac{1 - R}{3}) P_{\balpha}^\perp,
\label{Joseform}
\end{equation}
a convenient expression to note, where $P_{\balpha}$ and 
$P_{\balpha}^\perp$ are the projection operators on to the directions 
parallel and orthogonal, respectively, to the vector $\balpha$ 
for that vacuum.
  
The matrix $\phivac$ (\ref{Phivac}) at any one of the vacua,
being the vacuum (classical) value of the framon field there, 
would be the equivalent of the vierbeins $e^a_\mu$ in gravity 
for that vacuum, transforming between the local frame (here 
$su(3)$ labelled by the index $a$) and the global frame (here 
$\widetilde{su}(3)$ labelled by the index $\tilde{a}$).  The 
fact that this matrix is non-unitary for any $A$ means that 
at any of the degenerate vacua, the local and global frames, 
as also in gravity, cannot both be orthonormal.  But, whereas
in gravity it is the global frame indexed by $a$ which is 
taken orthonormal while the local frame indexed by $\mu$ is 
not, here on the other hand, it would be the other way round.  
Since $su(3)$ colour is supposed to be confining and exact, we 
would want the local frame to remain orthonormal, and it would 
be the global $\widetilde{su}(3)$ frame which is distorted.  
And just as in gravity, any vector or tensor quantity can be
given in either frame, i.e., labelled either by the global or
the local indices, or even partly by one and partly by the 
other.  The transformation between quantities carrying indices
of one type and quantities carrying indices of the other type  
can be effected simply by multiplying where appropriate with the 
matrix (\ref{Phivac}) or its conjugate, similar to the raising and
lowering of indices by the metric in gravity.  There will be examples 
later where such switches between local and global indices are 
found to be convenient.

\section{The Higgs States}

Next, from the framon potential $V[\Phi]$ in (\ref{VPhi}), one 
can deduce the spectrum of the Higgs states.  By Higgs states, we
mean, as usual, the quanta of fluctuations of the scalar fields,
i.e., in our case the framons, about the chosen vacuum, when these 
fluctuations do not correspond to the local gauge transformations
under which the theory is by construction invariant.  In the FSM, 
there are then two types of Higgs states.  First there is the 
ordinary or ``weak'' Higgs state coming from the fluctuations of 
the weak framon about the weak vacuum; this is the same as in 
the standard electroweak theory.  Secondly, there are 
the ``strong'' Higgs states, 
which come from the fluctuations of the strong framons about the 
strong vacuum, which we have now to identify for use in 
subsequent calculations.  

In the confinement interpretation \cite{tHooft,Bankovici} of 
symmetry-breaking that we find convenient to adopt, as explained 
in \cite{prepsm,efgt}, the Higgs states appear as bound states 
of framon-antiframon pairs, confined by ``weak'' $su(2)$ for the 
usual (weak) Higgs and by colour $su(3)$ for the strong Higgs states.  
In other words, the ``weak'' confinement being supposedly much 
deeper than colour confinement, the usual Higgs state will appear to 
us as fundamental while the strong Higgs states will appear to us 
as hadrons in what can be called the standard model 
scenario of confinement, where we can see only $su(2)$ singlets but
where we have already probed inside $su(3)$ singlets and coloured
objects are revealed. 

The strong vacuum having been elucidated in the preceding section,
it is in principle a straightforward matter to expand the framon
fields about the vacuum and identify those fluctuations which do
not correspond to gauge transformations.  For example, to exclude
those fluctuations corresponding to gauge transformations we can
work in a fixed gauge, say the hermitian gauge where $\Phi$ is
required to remain hermitian both before and after the fluctuations.  
Again, to avoid the complicated algebra when working directly with a 
general vacuum and non-orthonormal frames \cite{prepsm}, we adopt 
the tactic of the last section and first identify the strong Higgs 
states for the reference vacuum where things are simple, and then 
deduce the same for the general vacuum by an $\widetilde{su}(3)$ 
transformation.
   
Let us then start with the vacuum $\phivac^0$ in the simple
form (\ref{Phivac0}) and consider fluctuations of the framons 
$\Phi$ about it which we can choose to express as
\begin{equation}
\phivac^0 + \delta \Phi = \phivac^0 \,(1 + \epsilon S),
\label{deltaPhi}
\end{equation}
where we recall that $\phivac^0$ was chosen to be hermitian.
If we take $S$ to be $i \lambda_K,\ K = 1,...,8$, $\lambda_K$
being the standard Gell-Mann matrices, then the fluctuation will
not remain hermitian and would not then correspond, according to
our stated criterion, to a Higgs mode.  Indeed, it would instead
generate an $\widetilde{su}(3)$ transformation taking the chosen
vacuum to a neighbouring one.  Equivalently, by writing
\begin{equation}
\phivac^0 \,(1 + i \epsilon \lambda_K) = (1 + i \epsilon \lambda_K)
[(1 - i \epsilon \lambda_K) \,\phivac^0\, (1 + i \epsilon \lambda_K)]
\label{deltaPhia}
\end{equation}
we can see that it can be considered
as a fluctuation from the neighbouring vacuum in the hermitian
gauge (the factor inside square brackets), operated by a local $su(3)$ 
gauge transformation from the left,
and will represent a component of the $\Phi$ scalar field which, 
in the popular language of symmetry-breaking, is to be eaten up 
by one of the gauge vector bosons to give it a mass; it will not
correspond to a Higgs mode.  

The Higgs modes are to be represented by the other fluctuations 
with $S$ hermitian, and for which we can take $S$ as $S = 1$ or 
$S = \lambda_K,\ K = 1,...,8$.  However, for easier comparison 
with earlier results in \cite{prepsm}, we choose instead to work 
with some linear combinations of the above and write our Higgs 
basis states as
\begin{eqnarray}
V_1^0 & = & \left( \begin{array}{ccc} 1 & 0 & 0 \\
                                    0 & 0 & 0 \\
                                    0 & 0 & 0 \end{array} 
                                    \right); \nonumber \\
V_2^0 & = & \left( \begin{array}{ccc} 0 & 0 & 0 \\
                                    0 & 1 & 0 \\
                                    0 & 0 & 0 \end{array} 
                                    \right); \nonumber \\  
V_3^0 & = & \left( \begin{array}{ccc} 0 & 0 & 0 \\
                                    0 & 0 & 0 \\
                                    0 & 0 & 1 \end{array} 
                                    \right); \nonumber \\
V_4^0 & = & \left( \begin{array}{ccc}
            0 & \sqrt{\frac{1+2R}{2+R}} e^{i \phi_1} & 0 \\
            \sqrt{\frac{1-R}{2+R}} e^{-i \phi_1} & 0 & 0 \\
            0 & 0 & 0 \end{array} \right); \nonumber \\
V_5^0 & = & \left( \begin{array}{ccc}
            0 & 0 & \sqrt{\frac{1+2R}{2+R}} e^{i \phi_2} \\
            0 & 0 & 0 \\
            \sqrt{\frac{1-R}{2+R}} e^{-i \phi_2} & 0 & 0
            \end{array} \right); \nonumber \\
V_6^0 & = & \left( \begin{array}{ccc}
            0 & 0 & 0 \\
            0 & 0 & \frac{1}{\sqrt{2}} e^{i \phi_3} \\
            0 & \frac{1}{\sqrt{2}} e^{-i \phi_3} & 0
            \end{array} \right); \nonumber \\
V_7^0 & = & i \left( \begin{array}{ccc}
            0 & \sqrt{\frac{1+2R}{2+R}} e^{i \phi_1} & 0 \\
            -\sqrt{\frac{1-R}{2+R}} e^{-i \phi_1} & 0 & 0 \\
            0 & 0 & 0 \end{array} \right); \nonumber \\  
V_8^0 & = & i \left( \begin{array}{ccc}
            0 & 0 & \sqrt{\frac{1+2R}{2+R}} e^{i \phi_2} \\
            0 & 0 & 0 \\
            -\sqrt{\frac{1-R}{2+R}} e^{-i \phi_2} & 0 & 0
            \end{array} \right); \nonumber \\
V_9^0 & = & i \left( \begin{array}{ccc}
            0 & 0 & 0 \\
            0 & 0 & \frac{1}{\sqrt{2}} e^{i \phi_3} \\
            0 & -\frac{1}{\sqrt{2}} e^{-i \phi_3} & 0
            \end{array} \right).
\label{VK0}
\end{eqnarray}
They are chosen to form an orthonormal set when considered as
vectors in a 9-dimensional space and represent the 9 independent
Higgs states about the reference vacuum $\phivac^0$ in the 
$\widetilde {su}(3)$ gauge where ${\balpha}$ is $(1,0,0)$ and 
in the local $su(3)$ hermitian gauge. 

In the calculation that follows, we shall need the Higgs state
taken at other vacua of the degenerate set in the same gauges.  
These we can obtain again just by applying the
appropriate $\widetilde{su}(3)$ transformation $A^{-1}$ from the 
right as was done for the vacuum to obtain $\phivac$.  Thus, 
explicitly
\begin{equation}
V_K = V_K^0 A^{-1}.
\label{VK}
\end{equation}
It is easy to check that orthonormality of these $V_K$ will be 
preserved, since the 9-dimensional inner product is given in terms of
the matrices as $\tr (V_K V_L^\dagger)$.

Notice that these 9 Higgs states form a complete orthonormal set
but they are not in general mass eigenstates.   Together with the
Higgs state from the electroweak sector, we have a $10 \times 10$
mass matrix for the Higgs states which can be computed by taking 
the second derivatives of the framon potential $V[\Phi]$, and 
this can then be diagonalized to find the eigenstates.  Again, 
for the strong sector it is easiest to calculate the Higgs mass
matrix first in the gauges where the vacuum (\ref{Phivac0}) is 
diagonal and then to transform it sandwiching with $A$ and $A^{-1}$ 
as necessary, since the eigenvalues will be unchanged by this 
transformation.  

The Higgs mass matrix and its diagonalization will not be
needed in this paper directed towards the understanding of 
the mass and mixing patterns of quarks and leptons, and so
will be relegated to Appendix A.  However, for testing the
FSM as a whole, the Higgs mass matrix may play in future a 
central role.  The main new ingredients introduced by the
FSM, we recall, are the strong framons which alone have no 
standard model counterparts, and the strong Higgs states
under present consideration are their direct manifestations.  
Hence, the spectroscopy of these strong Higgs states would
seem deserving of a close scrutiny when more is known about 
the parameters which enter into the model.

One consequence of the Higgs mass matrix (\ref{higgsmassmatrix}) in 
Appendix 
A which could be of immediate phenomenological interest is 
the fact that two of the strong Higgs states have the same 
quantum numbers as $h$, the ``weak '' or standard model 
Higgs state, and so can mix with the latter.  And these 
strong Higgs states being hadrons, with presumably hadronic 
decay widths and modes, any admixture of them into $h$ can 
greatly alter the latter's decay characteristics deduced 
from the standard model and presently used experimentally as 
signatures for its detection.  This possibility is under 
study. 

\section{The Scale-Dependence of the Vacuum}

Next, we are to study the change in the (strong) vacuum with 
changing scales under renormalization by loops of the (strong) 
Higgs states, so as to derive its effects, if any, on the rotation 
of the quark and lepton mass matrix.  We recall
that since the (strong) vacuum is coupled to the quark and 
lepton mass matrix (\ref{mfact}) via the vector $\balpha$ which
appears in the $\nu_2$ term of the framon potential $V[\Phi]$,
any change in it will get reflected in the
quark and lepton mass matrix (\ref{mfact}).  To get $\balpha$,
a vector in $\widetilde{su}(3)$, to rotate with scale, we would
want renormalization effects which are not $\widetilde{su}(3)$ 
invariant.  And since the (strong) vacuum breaks this symmetry,
as shown above, so also will the (strong) Higgs states derived
as fluctuations about this vacuum. The renormalization effects
from these will thus satisfy the above criterion, hence our
interest.

Information on the effects on the vacuum under renormalization 
can in principle be derived through any quantity which depends 
on the vacuum and gets renormalized by the strong Higgs loops.  
We choose, mainly for historical reasons, to focus on the 
self-energy of certain fermion states to be specified on which 
we had some experience earlier in a similar context \cite{ckm}.  
As will be seen, the information from this study is enough 
already to show that rotation will result.  A parallel study on 
other quantities which we have not performed can in principle 
give further constraints on the scale-dependence of the vacuum, 
and hence on the rotation of $\balpha$, but these should not be 
in contradiction with what we have derived if the present theory 
is self-consistent.

To specify these fermion states, 
let us first remind ourselves of the Yukawa term in 
the FSM for the weak framon written down in \cite{
efgt,prepsm} from which the mass matrix (\ref{mfact}) is derived:
\begin{eqnarray}
{\cal A}^{\rm lepton}_{\rm YK} &=& \sum_{[\tilde{a}] [b]} Y_{[b]} 
\bar{\psi}^r_{[\tilde{a}]}
    \phi_{r}^{(-) \tilde{a}} \half (1 + \gamma_5) \psi^{[b]}
    + \sum_{[\tilde{a}] [b]} Y'_{[b]} \bar{\psi}^r_{[\tilde{a}]}
    \phi_{r}^{(+) \tilde{a}} \half (1 + \gamma_5) \psi'^{[b]}
    \nonumber \\
&& {} + {\rm h.c.}
\label{Yukawaw}
\end{eqnarray}
for leptons (similarly for quarks).  The mass matrix is 
obtained by substituting, for the (weak) framon field, its 
vacuum value; then by a suitable relabelling of the right-handed
fields, as indicated in the introduction, the mass matrix can 
be recast in the factorizable form (\ref{mfact}).  Furthermore,
by expanding the framon field about its vacuum value to first 
order, one obtains the Yukawa coupling of the (weak) Higgs state
$h$ to leptons.  In the confinement picture both the leptons
and the Higgs state are bound states via $su(2)$ confinement,
the former of a (weak) framon with a fundamental fermion $\psi$
and the latter of a (weak) framon-antiframon pair.  Similar
assertions, of course, can be made about quarks.

Our object now is to write down a similar Yukawa term for 
the strong framon, with colour $su(3)$ now taking the place 
of the electroweak $su(2)$ in (\ref{Yukawaw}).  For this, the 
following expression was suggested \cite{prepsm}
\begin{equation}
{\cal A}^{\rm strong}_{\rm YK} = \sum_{[b]} Z_{[b]} \bar{\psi}^a \bphi_a
\cdot \balpha_0
    \frac{1}{2}(1 + \gamma_5) \psi^{[b]} + {\rm h.c.}
\label{Yukawas}
\end{equation}

We note here that in (\ref{Yukawas}) the fermion fields do 
not carry any $\tilde{a}$ index for $\widetilde{su}(3)$ but 
the framon fields $\bphi_a$ do since they are vectors in generation space.  
Hence, to maintain $\widetilde{su}(3)$ 
invariance, we need a vector in $\widetilde{su}(3)$ space to 
saturate this $\tilde{a}$ index.  There is no such vector 
available to play this role within the purely strong sector, 
but in the present FSM set-up, there is the vector $\balpha$ 
coming from the weak sector which can be so employed.  In 
introducing here a vector originating from the weak sector 
to construct the Yukawa term (\ref{Yukawas}) for the strong 
framon in the strong sector, one is imitating, in spirit though 
not in detail, the construction of the standard Yukawa term 
(\ref{Yukawaw}) in the weak sector.  In fact, as it stands, the weak
Yukawa term in (\ref{Yukawaw}) is not explicitly invariant
under $\widetilde{su}(2)$ as it ought to be, but it can be 
put in an explicitly invariant form \cite{efgt}  by writing   
 $\phi_r^{(\pm) \tilde{a}}$ as $\bgamma^{(\pm)}
\cdot\bphi_r^{\tilde{a}}$ in terms of the $\widetilde{su}(2)$ 
vectors $\bgamma^{(\pm)}$ originating from the electromagnetic 
$u(1)$ sector.  This is in close parallel to the introduction 
above of the vector $\balpha$ originating from the weak sector 
to keep (\ref{Yukawas}) $\widetilde{su}(3)$ invariant.   The 
vector $\balpha$ here, however, does not have a definite value, 
but can point in any direction in $\widetilde{su}(3)$ space 
since the vacuum is degenerate.  Nevertheless, these directions 
being all gauge equivalent, it should not matter which value we 
choose.  By writing in (\ref{Yukawas}) ${\balpha}_0$ for the 
vector ${\balpha}$, we have implicitly chosen that value for 
$\balpha$ which corresponds to the reference vacuum in  
\S 2 above, or conversely that we have chosen the reference vacuum 
in \S 2 to be the vacuum corresponding to that vector 
$\balpha$ appearing in the Yukawa coupling (\ref{Yukawas}).  
The physical meaning for $\balpha_0$ will be apparent later.

The fermion field $\psi$ appearing in (\ref{Yukawas}) above
was originally meant \cite{prepsm} to be only generic, since 
it was thought that for studying renormalization effects on 
the vacuum it ought not to matter with which fermion field 
one started.  The question arises, however, in the realistic 
situation, whether any fermion field exists which is of the 
generic type $\psi$ that one wants.  Now, for reasons one 
does not yet understand, the standard model admits only 
$su(2)$ doublet left-handed fermions and only $su(2)$ singlet 
right-handed fermions.  In that case, the $\psi$ field in 
(\ref{Yukawas}) if interpreted as a fundamental field would 
appear to go against the grain, being an $su(2)$ singlet and 
left-handed.  We recall, however, that in the confinement 
picture we adopted, quarks (and leptons) appear as bound 
states of a fundamental fermion field with a weak framon via 
$su(2)$ confinement.  Hence, starting with the left-handed 
$su(2)$ doublet fundamental fermion field $\psi_r^{[\tilde{a}]}$ 
appearing already in (\ref{Yukawaw}) and the weak framon field 
$\phi_r^{\tilde{r} \tilde{a}}$ in (\ref{wframon}), one easily 
obtains a quark field
\begin{equation}
\psi^{\tilde{r}} = \sum_{r,\tilde{a}} 
   \phi_r^{\tilde{r} \tilde{a}} \psi_r^{[\tilde{a}]},
\label{psirtilde}
\end{equation}
which is a left-handed $su(2)$ singlet as required.  It does
carry an $\widetilde{su}(2)$ index $\tilde{r}$ but this is a 
global index which can be saturated in (\ref{Yukawas}) with 
the $\tilde{r}$ index in the factor $\beta^{\tilde{r}}$ 
originally carried by the strong framon (\ref{sframon}), only
suppressed in (\ref{Yukawas}) for convenience.  In other words, 
there are indeed $\psi$ fields of the generic type required in 
(\ref{Yukawas}), only to be interpreted as quark fields, not 
as fundamental fields.  Hence, for studying the effects of
renormalization on the strong vacuum that we are after, the
Yukawa term (\ref{Yukawas}) is indeed admissible, though not 
at the fundamental level, but in the standard model scenario 
of interest to us.

Starting then from the Yukawa term (\ref{Yukawas}), now so 
interpreted, one can proceed, as one did in (\ref{Yukawaw}) 
above, by inserting for the framons their vacuum values to
derive a mass matrix, thus
\begin{equation}
\bm= \zeta_S |v_0 \rangle \langle Z| \half (1+\gamma_5) + \zeta_S
|Z\rangle \langle v_0 | \half (1-\gamma_5),
\label{mwithgamma}
\end{equation}
where $\langle Z| = (Z_{[1]}, Z_{[2]}, Z_{[3]})$.  We can 
make $\bm$ hermitian and independent of $\gamma_5$ as we 
did for (\ref{mfact}), following Weinberg \cite{Weinbergren}, by relabelling
the right-handed fields to obtain the form
\begin{equation}
{\bf m} = {\bf m}_T |v_0 \rangle \langle v_0|,
\label{mfacts}
\end{equation}
with
\begin{equation}
|v_0 \rangle = V_0 {\balpha}_0
\label{v0ket}
\end{equation}
and
\begin{equation}
{\bf m}_T = \zeta_S \rho_S/v_0,\ \ 
   \rho_S = \sqrt{Z_{[1]}^2 + Z_{[2]}^2 + Z_{[3]}^2},\ \ 
   v_0 = \sqrt{\langle v_0|v_0 \rangle}.
\label{bfmT}
\end{equation}
This ${\bf m}$ then is the mass matrix for the fermions the 
self-energy of which we propose to study under renormalization.  

We note that these fermions bear the same relationship to 
the strong framons in (\ref{Yukawas}) as did the leptons 
to the weak framons in (\ref{Yukawaw}).  Hence if, in the 
confinement picture of 't~Hooft \cite{tHooft} and of Banks
and Rabinovici \cite{Bankovici}, one interprets the leptons 
as bound states of the fundamental fermion $\psi$ with the 
weak framon $\phi$ via weak $su(2)$ confinement, then the 
present fermions should be thought of as bound states of 
the fermion field $\psi$ (quarks) with the strong framon 
$\Phi$ via strong $su(3)$ (i.e., colour) confinement.  In 
other words, they are to be interpreted as hadrons, and as 
such will interact strongly with the strong Higgs states
listed in the preceding section which are likewise hadrons.

As given in (\ref{mfacts}), both the rows and columns of 
the matrix $\bm$ are labelled by colour $su(3)$ indices, 
since $|v_0 \rangle$ according to (\ref{v0ket}) is a vector 
in $su(3)$ space ($V_0$ being a matrix with rows labelled 
by $su(3)$ but columns by $\widetilde{su}(3)$ indices, and 
 $\balpha_0$  a vector in $\widetilde{su}(3)$ 
space).  This may seem a little surprising when considered
as the mass matrix of the bound states just mentioned, of
the fields $\psi$ and $\Phi$ by colour confinement, thus
\begin{equation}
\bchi = \Phi^\dagger \bpsi \sim \phivac^\dagger \bpsi
\label{bchi}
\end{equation}
which ought to be indexed by $\widetilde{su}(3)$ indices, 
not by colour indices which are saturated in (\ref{bchi}), 
as they should be, colour being confined.  However, one
notes that the conversion from $\bpsi$ to $\bchi$ is only
via multiplication by the matrix $\phivac$ which, as
noted before at the end of \S3, plays here the role 
of the vierbeins $e^a_\mu$ in gravity.  So just as, say, 
the Ricci tensor in gravity can be represented either as
$R^{ab}$ or as $R_{\mu\nu}$, the two expressions being
related by contracting with the vierbeins $e^a_\mu$, so
the mass matrix $\bm$ here can be represented either as 
as a matrix labelled by the local index $a$ or the global
index $\tilde{a}$, the two expressions being related 
by the matrix $\phivac$.  It just so happens that for
our purpose here, it is more convenient to work with the
version (\ref{mfacts}) above labelled by the local $su(3)$
indices.    

Next, to examine how this mass matrix ${\bf m}$ renormalizes
through insertions of strong Higgs loops, we
shall need the couplings of these fermions with the strong
Higgs states.  To derive these couplings, we follow the same
procedure as in (\ref{deltaPhi}) above and expand now to first
order the strong framon field $\Phi$ in (\ref{Yukawas}) about 
its vacuum value, thus
\begin{equation}
\Phi \sim \phivac + \sum_K H_K V_K,
\label{Phiexp}
\end{equation}
with $V_K = V_K^0 A^{-1}$, $V_K^0$ being any one of the 9 
matrices listed in (\ref{VK0}) above, and $H_K$ one of the 
strong Higgs fields.  Substituting this into (\ref{Yukawas})
and recalling that we have already relabelled there the 
right-handed fermion fields so as to derive the mass matrix
${\bf m}$ in the form (\ref{mfacts}), one easily obtains the 
desired couplings as
\begin{equation}
\Gamma_K = \rho_S |v_K \rangle \langle v_0| \half
   (1 + \gamma_5) + \rho_S |v_0 \rangle \langle v_K| 
   \half (1 - \gamma_5),
\label{GammaK}
\end{equation}
with
\begin{equation}
|v_K \rangle = V_K {\balpha}_0.
\label{vKket}
\end{equation}

With these couplings $\Gamma_K$, we can now evaluate the 
insertion of a strong Higgs loop to the fermions self-energy 
as
\begin{equation}
\Sigma(p) = \frac{i}{(4 \pi)^4} \sum_K \int d^4 k \frac{1}{k^2 
   - M_K^2} \Gamma_K \frac{(p\llap/ - k\llap/) + \bm}{(p - k)^2 
   - {\bm}^2} \Gamma_K,
\label{Sigma}
\end{equation}
where we may for the moment take $K$ to label the Higgs mass 
eigenstates.  After standard manipulations and regularizing the 
divergence by dimensional regularization, one obtains: 
\begin{equation}
\Sigma(p) = - \frac{1}{16 \pi^2} \sum_K \int_0^1 dx \Gamma_K
    \{\bar{C} + \ln (\mu^2/Q^2) \} [p\llap/ (1 - x) + \bm]\, \Gamma_K,
\label{Sigma2}
\end{equation}
where
\begin{equation}
Q^2 = \bm^2 x + M_K^2 (1 - x) - p^2 x(1 -x),
\label{Qsquare}
\end{equation}
with $\bar{C}$ being the divergent constant to be subtracted in the 
standard $\overline{\rm MS}$ scheme.  The change to the mass matrix under renormalization, $\delta {\bf m}$, is obtained by first commuting the $p\llap/$ 
in the numerator half to the extreme left and half to the extreme 
right, then putting $p\llap/ = {\bf m}$ and $p^2 = {\bf m}^2$.  The 
full explicit expression for $\delta \bm$ so obtained together with 
more details of the calculation can be found in \cite{transmudsm}.  
Here, we shall interest ourselves only in the terms proportional to
$\ln \mu^2$ and hence dependent on the scale $\mu$.  These are of 
two types.  First, there are terms of the form
\begin{equation}
\Gamma_K \bm \Gamma_K = \rho_S^2 
    \langle v_0|v_K \rangle |v_K \rangle \langle v_0|
    \half (1 + \gamma_5) + {\rm c.c.}
\label{GammamGamma}
\end{equation}
Then there are terms of the form
\begin{eqnarray}
\lefteqn{ \Gamma_K p\llap/ \Gamma_K  \rightarrow} \nonumber \\ 
&&\half\,\rho_S^2
    \{ \langle v_K|v_0 \rangle |v_K \rangle
    \langle v_0| +  \langle v_K|v_K \rangle
    |v_0 \rangle \langle v_0| \} \half (1 + \gamma_5) + \ {\rm c.c.}
\label{GammapGamma}
\end{eqnarray}
In (\ref{GammamGamma}) and (\ref{GammapGamma}), we have already 
commuted $p\llap/$ to the left and right as stipulated and used 
the known forms for the tree-level mass matrix (\ref{mfacts}) and 
Higgs couplings (\ref{GammaK}).

This then gives the renormalized mass matrix ${\bf m}'$ as
\begin{equation}
\bm'= \zeta_S |\tilde{v}_0 \rangle \langle v_0| \half (1+\gamma_5) 
   + \zeta_S |v_0 \rangle \langle \tilde{v}_0 | \half (1-\gamma_5),
\label{remasswithgamma}
\end{equation}
with
\begin{equation}
|\tilde{v}_0 \rangle = |v_0 \rangle + (\ln \mu^2)\,a |v_0 \rangle + 
 (\ln \mu^2)\,|u \rangle,
\label{v0p}
\end{equation}
where
\begin{equation}
a = - \frac {1}{16 \pi^2} \rho_S^2 \left( \frac{5}{4}
   + \frac{1}{4} \sum_{K=1,\ldots,9} \langle v_K|v_K \rangle \right),
\label{a}
\end{equation}
and
\begin{equation}
|u \rangle = -\frac {1}{16 \pi^2} \rho_S^2 \left(\frac{3}{4}\right)
               \sum_{K=7,8,9}\langle v_0| v_K\rangle |v_K \rangle,
\label{v1}
\end{equation}
where we notice that, both $\bm$ and $\Gamma_K$ being factorizable,
the renormalized $\bm'$ remains also factorizable.  Again, $\bm'$ 
can made hermitian and without $\gamma_5$, i.e., into the same form  
as $\bm$ in (\ref{mfacts}) by relabelling the right-handed fields; 
so what has changed by renormalization is really just the left-hand 
factor $\zeta_S |v_0 \rangle$, for which we shall denote temporarily 
as $|w \rangle$, so that we have from (\ref{v0p})
\begin{equation} 
|w' \rangle = |w \rangle + (\ln \mu^2)\,a |w \rangle + 
 (\ln \mu^2)\,\zeta_S |u \rangle
\label{wprime},
\end{equation}
The quantity $|w'\rangle$ is the
value of $|w\rangle$ at $\mu + \delta \mu$, so that 
in the limit as $\delta \mu \to 0$, we get
\begin{equation}
\frac{d}{d \ln \mu^2} |w\rangle =
  \frac{d}{d \ln \mu^2} \left( \zeta_S |v_0 \rangle \right) 
   = a \zeta_S |v_0 \rangle + \zeta_S |u \rangle,
\label{RGE}
\end{equation}
where now both $\zeta_S$ and $|v_0\rangle$ are considered as varying
with respect to the scale $\mu$.

Recalling from (\ref{v0ket}) above that $|v_0 \rangle = V_0 \balpha_0$
where $V_0 = \phivac$ represents the vacuum value of the strong
framon field $\Phi$, we deduce the fact that $|v_0 \rangle$ varies 
with $\mu$ means that the vacuum will vary with $\mu$ also, as 
anticipated.  And since, according to (\ref{Phivac}), $\phivac$ 
is given by an $\widetilde{su}(3)$ transformation on the 
reference vacuum $\phivac^0$ in the chosen gauges where the 
latter is diagonal, the change of $\phivac$ with respect to scale 
can be transferred to the change of the matrix $A$ representing that 
transformation.

To exhibit explicitly how $A$ will depend on scale according to the
equation (\ref{RGE}) derived above, let us write now
\begin{equation}
A = R_1 R_2 R_3 P,
\label{A3G}
\end{equation}
with
\begin{eqnarray}
R_1 & = & \left( \begin{array}{ccc} 1 & 0 & 0 \\
        0 & c_1 & -s_1 e^{-i \sigma_1} \\ 0 & s_1 e^{i \sigma_1} & c_1
        \end{array} \right) \nonumber \\
R_2 & = & \left( \begin{array}{ccc} c_2 & 0 & -s_2 e^{-i \sigma_2} \\
        0 & 1 & 0 \\ s_2 e^{i \sigma_2} & 0 & c_2 \end{array} \right)
        \nonumber \\
R_3 & = & \left( \begin{array}{ccc} c_3 & -s_3 e^{-i \sigma_3} & 0 \\
        s_3 e^{i \sigma_3} & c_3 & 0 \\ 0 & 0 & 1 \end{array} \right)
        \nonumber \\
P & = & \left( \begin{array}{ccc} e^{i \alpha_1} & 0 & \\
      0 & e^{i \alpha_2} & 0 \\ 0 & 0 & e^{-i \alpha_1 -i \alpha_2}
      \end{array} \right),
\label{A3Gd}
\end{eqnarray}
where $c_i = \cos \theta_i, s_i = \sin \theta_i$.  The expression 
(\ref{A3G}) is the standard parametrization in terms of Euler angles 
but deliberately taken in reverse order for a reason which will soon 
be apparent.

For the same reason, we choose to rewrite $A^{-1}$ as
\begin{equation}
A^{-1} = P_1^{-1} R_3^{'-1} R_2^{'-1} R_1^{'-1},
\label{Ainv3G}
\end{equation}
where
\begin{equation}
P_1^{-1} = P^{-1} P_2, R_3^{'-1} = P_2^{-1} R_3^{-1} P_2, 
R_2^{'-1} = P_2^{-1} R_2^{-1} P_2, R_1^{'-1} = P_2^{-1} R_1^{-1},
\label{Ainv3Gd}
\end{equation}
with
\begin{equation}
P_2 = \left( \begin{array}{ccc} 1 & 0 & 0 \\
      0 & e^{i \alpha_2} & 0 \\ 0 & 0 & e^{-i \alpha_2} \end{array}
      \right).
\label{P1P2}
\end{equation}
Explicitly
\begin{eqnarray}
P_1^{-1} & = & \left( \begin{array}{ccc} e^{-i \alpha_1} & 0 & 0 \\
   0 & 1 & 0 \\ 0 & 0 & e^{i \alpha_1} \end{array} \right)
   \nonumber \\
R_3^{'-1} & = & \left( \begin{array}{ccc} c_3 & s_3 e^{-i \sigma'_3} 
   & 0 \\ - s_3 e^{i \sigma'_3} & c_3 & 0 \\ 0 & 0 & 1 \end{array}
   \right) \nonumber \\
R_2^{'-1} & = & \left( \begin{array}{ccc} c_2 & 0 & s_2 e^{-i 
   \sigma'_2} \\ 0 & 1 & 0 \\ - s_2 e^{i \sigma'_2} & 0 & c_2 
   \end{array} \right) \nonumber \\
R_1^{'-1} & = & \left( \begin{array}{ccc} 1 & 0 & 0 \\
   0 & c_1 e^{-i \alpha_2} & s_1 e^{-i \sigma'_1} \\
   0 & -s_1 e^{i \sigma'_1} & c_1 e^{i \alpha_2} \end{array} \right),
\label{Ainv3Gdp}
\end{eqnarray}
where
\begin{equation}
\sigma'_3 = \sigma_3 - \alpha_2, \ \ \sigma'_2 = \sigma_2 + \alpha_2,
   \ \ \sigma'_1 = \sigma_1 + \alpha_2,
\label{sigmaip}
\end{equation}
and $R'_1$ is a general element of the $\widetilde{su}(2)$ little
group which leaves ${\balpha}_0$ invariant.  

The reason we choose to write $A^{-1}$ in this way is that $V_0$, 
in which $A$ appears, enters into the rotation equations only as
$|v_0 \rangle = V_0 {\balpha}_0$, so that, given the invariance of
${\balpha}_0$ under $R'_1$, we have
\begin{equation}
|v_0 \rangle = V_0 {\balpha}_0 =\phivac^0 P_1^{-1} R_3^{'-1} R_2^{'-1} 
{\balpha}_0.
\label{v0ket3G}
\end{equation}
This means that $|v_0 \rangle$ and hence also the rotation equations  
are actually independent of the variables $\theta _1, \sigma'_1$ and 
$\alpha_2$, the parameters of  $R'_1$.  Thus, of the 8 original 
parameters in $A$ as befits an  $\widetilde{su}(3)$ transformation, 
there remain only the following 5 which figure in (\ref{RGE}), namely
$\theta_2, \theta_3, \alpha_1, \sigma'_2, \sigma'_3$.  Writing $A^{-1}$ 
in the form (\ref{Ainv3G}) thus allows one to remove the unnecessary 
variables from the equation and gives an expression of $|v_0 \rangle$ 
in only the remaining 5 on which it really depends.

\section{The Rotation Equation: RGE for ${\balpha}$}

Substitution of (\ref{Ainv3G}) for $A^{-1}$ into (\ref{RGE}) then
gives equations for the 5 variables on which $A$ depends, which is
all the information one needs.  However, what interests us in the 
end is actually the vector
\begin{equation}
{\balpha} = A {\balpha}_0,
\label{alpha0}
\end{equation}
appearing in the mass matrix $m$ in (\ref{mfact}) for leptons
and quarks, which we may parametrize in general as
\begin{equation}
{\balpha} = \left( \begin{array}{l} 
                            \cos \theta \,e^{-i \beta_1} \\
                            \sin \theta \sin \phi \,e^{-i \beta_2} \\
                            \sin \theta \cos \phi \,e^{-i \beta_3}
                            \end{array} \right).
\label{alpha3G}
\end{equation}
It pays therefore to express $|v_0 \rangle$, and hence subsequently 
also the rotation equations directly, not in terms of the variables
$\theta_2, \theta_3, \alpha_1, \sigma'_2, \sigma'_3$ above but in 
terms of the parameters of ${\balpha}$.  This we can do by working 
out (\ref{alpha0}) in terms of these parameters appearing in $A$ 
above and comparing with (\ref{alpha3G}), then solving for one set 
of variables in terms of the other.  This then allows us to write:
\begin{equation}
|v_0 \rangle = \left( \begin{array}{l} 
   \sqrt{\frac{1+2R}{3}} \cos \theta \,e^{i \beta_1} \\
   - \sqrt{\frac{1-R}{3}} \frac{\sin \theta \cos \theta \sin \phi}
     {\sqrt{\cos^2 \theta + \sin^2 \theta \cos^2 \phi}}
     \,e^{-i \beta_2 + i \beta_1} \\
   - \sqrt{\frac{1-R}{3}} \frac{\sin \theta \cos \phi}
     {\sqrt{\cos^2 \theta + \sin^2 \theta \cos^2 \phi}}
     \,e^{-i \beta_3} \end{array} \right),
\label{v03G}
\end{equation}
with norm given by
\begin{equation}
v_0^2=\langle v_0|v_0 \rangle = \frac{1}{3} (1 + 2R \cos^2 \theta 
   -R \sin^2 \theta),
\label{v0norm3G}
\end{equation}
an expression we shall need later.

Substituting (\ref{v03G}) into (\ref{RGE}) now gives an equation
in the desired variables, namely $R, \theta, \phi$ plus the 3
phases $\beta_1, \beta_2, \beta_3$, where the left-hand side can
be written as
\begin{equation}
\frac{d}{d \ln \mu^2} |w\rangle = \dot{\zeta}_S |w_{\zeta} \rangle 
+ \dot{\theta} |w_{\theta} \rangle + \dot{\phi} |w_{\phi}
\rangle +  \sum_{i=1,2,3} \dot{\beta_i} |w_{\beta_i} \rangle,
\label{zetav0dot}
\end{equation}
where a dot denotes differentiation with respect to $\ln \mu^2$ and where
\begin{equation}
|w_{\zeta} \rangle  = \frac{\partial}{\partial \zeta_S} 
    |w \rangle,\ \  \rm{etc.}
\label{v0zeta}
\end{equation}
Differentiating then first with respect to the phases, we get
\begin{eqnarray}
|w_{\beta_1} \rangle & = & i \zeta_S \left( \begin{array}{l}
\sqrt{\frac{1+2R}{3}} \cos \theta \,e^{i \beta_1} \\
   - \sqrt{\frac{1-R}{3}} \frac{\sin \theta \cos \theta \sin \phi}
     {\sqrt{\cos^2 \theta + \sin^2 \theta \cos^2 \phi}}
     \,e^{-i \beta_2 + i \beta_1} \\
 0 \end{array} \right) \label{wbeta1}\\
|w_{\beta_2} \rangle & = &i \zeta_S \left( \begin{array}{l}
0\\ 
\sqrt{\frac{1-R}{3}} \frac{\sin \theta \cos \theta \sin \phi}
     {\sqrt{\cos^2 \theta + \sin^2 \theta \cos^2 \phi}}
     \,e^{-i \beta_2 + i \beta_1} \\
 0 \end{array} \right) \label{wbeta2}\\
|w_{\beta_3} \rangle & = & i \zeta_S \left( \begin{array}{l}
0\\ 0 \\
\sqrt{\frac{1-R}{3}} \frac{\sin \theta \cos \phi}
     {\sqrt{\cos^2 \theta + \sin^2 \theta \cos^2 \phi}}
     \,e^{-i \beta_3} \end{array} \right) \label{wbeta3},
\end{eqnarray}
where we notice that apart from the equal phases on each of the 
3 components on both sides, the right-hand sides of all these 
equations are all imaginary, whereas the left-hand side of 
(\ref{zetav0dot}) is real.  Hence we conclude that 
\begin{equation}
\dot{\beta}_1 =\dot{\beta}_2 = \dot{\beta}_3 = 0.
\label{betadots}
\end{equation}

The remaining three partial derivatives give
\begin{eqnarray}
|w_{\zeta} \rangle &=& \frac{1}{\sqrt{3}} \left( \begin{array}{c}
   \frac{1}{\sqrt{1+2R}} \cos \theta \,e^{i \beta_1} \\
   - \frac{1}{\sqrt{1-R}} \frac{\sin \theta \cos \theta \sin \phi}
     {\sqrt{\cos^2 \theta + \sin^2 \theta \cos^2 \phi}}
     \,e^{-i \beta_2 + i \beta_1} \\
   - \frac{1}{\sqrt{1-R}} \frac{\sin \theta \cos \phi}
     {\sqrt{\cos^2 \theta + \sin^2 \theta \cos^2 \phi}} 
     \,e^{-i \beta_3} \end{array} \right) \nonumber \\
|w_{\theta} \rangle & = & \frac{\zeta_S}{\sqrt{3}} \left(
   \begin{array}{c} - \sqrt{1+2R} \sin \theta \,e^{i \beta_1} \\
   - \sqrt{1-R} \frac{\sin \phi(\cos^4 \theta - \sin^4 \theta
   \cos^2 \phi)}{(\cos^2 \theta + \sin^2 \theta \cos^2 \phi)^{3/2}}
   \,e^{-i \beta_2 + i \beta_1} \\
   - \sqrt{1-R} \frac{\cos \theta \cos \phi}
   {(\cos^2 \theta + \sin^2 \theta \cos^2 \phi)^{3/2}} 
   \,e^{-i \beta_3} \end{array} \right) \nonumber \\
|w_{\phi} \rangle & = & \frac{\zeta_S}{\sqrt{3}} \left(
   \begin{array}{c} 0 \\ - \sqrt{1-R} \frac{\sin \theta \cos \theta
     \cos \phi}{(\cos^2 \theta + \sin^2 \theta \cos^2 \phi)^{3/2}}
     \,e^{-i \beta_2 + i \beta_1} \\
   \sqrt{1-R} \frac{\sin \theta \cos^2 \theta \sin \phi}
     {(\cos^2 \theta + \sin^2 \theta \cos^2 \phi)^{3/2}} 
     \,e^{-i \beta_3} \end{array} \right).
\label{v0zeta3G}
\end{eqnarray}
In order to make comparison with (\ref{RGE}), we have to compute
the following quantities:
\begin{equation}
\sum_K \langle v_K|v_K \rangle = \frac{3}{2+R}(1 + E)
\label{SvK3G}
\end{equation}
with
\begin{equation}
E = 1 - R \cos^2 \theta + 2 R \sin^2 \theta,
\label{E3G}
\end{equation}
and
\begin{equation}
\sum_{K=7,8,9}\langle v_0| v_K\rangle |v_K \rangle
   = - \frac{3 R \sin \theta \cos \theta}{2+R} \left( 
     \begin{array}{l} \sqrt{\frac{1+2R}{3}} \sin \theta \,e^{i \beta_1} \\
     \sqrt{\frac{1-R}{3}} \frac{\cos^2 \theta \sin \phi}
       {\sqrt{\cos^2 \theta + \sin^2 \theta \cos^2 \phi}}
         \,e^{i \beta_1 - i \beta_2 } \\
     \sqrt{\frac{1-R}{3}} \frac{\cos \theta \cos \phi}
       {\sqrt{\cos^2 \theta + \sin^2 \theta \cos^2 \phi}}
         \,e^{-i \beta_3} \end{array} 
     \right).
\label{gvI3G}
\end{equation}
Notice that in computing these quantities, the state vectors
$|v_K \rangle$ are summed over in a manner invariant under an 
orthogonal transformation among the Higgs states $V_K$, so 
that these Higgs states need not be taken as the actual mass 
eigenstates they started out to be in (\ref{Sigma}) but can 
be taken as any convenient orthonormal set, in particular just
those $V_K = V_K^0 A^{-1}$ for $V_K^0$ listed in (\ref{VK0}).
Notice also that the phases in (\ref{v0zeta3G}), (\ref{gvI3G})
are the same as in $|v_0 \rangle$.  With these, equation 
(\ref{RGE}) is now made explicit.

We can then extract from (\ref{RGE}) the desired equations for 
$\dot{\zeta}_S, \dot{\theta}, \dot{\phi}$ by taking the inner 
products with the vectors $|w_{\zeta} \rangle,|w_{\theta} \rangle, 
|w_{\phi} \rangle$, obtaining
\begin{eqnarray}
\dot{R} &=& \frac{\rho_S^2}{16 \pi^2} \frac{R(1+2R)(1-R)}{E}
   \left[\frac{5}{2} + \frac{1}{2} \frac{3}{2+R} (1+E) \right],
\label{Rdot}\\
\dot{\theta} &=& \frac{\rho_S^2}{16 \pi^2} \frac{R \sin 2 \theta}{E}
   \left(\frac{3}{2}\right) \left[\frac{5}{4} + \frac{3}{4} \frac{1}{2+R} 
   \right],
\label{thetadot}\\
\dot{\phi} &=& \frac{\rho_S^2}{16 \pi^2} \frac{R \sin^2 \theta
   \sin 2 \phi}{E} \left(\frac{3}{2}\right) \left[\frac{5}{4} + \frac{3}{4}
   \frac{1}{2+R} \right],
\label{phidot}
\end{eqnarray}
where the last 2 imply that
\begin{equation}
\dot{\phi}/\dot{\theta} = \half \tan \theta \sin 2 \phi,
\label{Joseeq1}
\end{equation}
which integrates to
\begin{equation}
\cos \theta \tan \phi = {\rm constant},
\label{Joseeq2}
\end{equation}
a condition one can use in place of (\ref{phidot}) above.

These equations (\ref{Rdot})---(\ref{phidot}), although derived 
only from a single (strong) Higgs loop insertion and therefore 
likely to be rather limited both in accuracy and in range of
applicability, show nevertheless the crucial fact that the vector 
$\balpha$, appearing in the mass matrix (\ref{mfact}) for quarks 
and leptons, does indeed rotate as anticipated.  The rotation is 
driven by the strong interactions and can thus be fast enough for 
the effects we want.  Specifically, in the equations 
(\ref{Rdot})---(\ref{phidot}), 
the speed of rotation is governed by the strong 
coupling $\rho_S$ between the hadron fermion state whose mass is 
being renormalized to the strong Higgs states, and this coupling 
can be adjusted to fit data if so desired.  Besides, the rotation 
originates from the scale-dependence of the vacuum and only gets
transmitted to the fermion mass matrix (\ref{mfact}) via this 
$\balpha$ by virtue of the appearance in the Yukawa coupling 
(\ref{Yukawaw}) of the weak framon field (\ref{wframon}) of which 
$\balpha$ is a factor.  The value of $\balpha$, and also the manner
it rotates, is thus independent of the whether it appears in the
mass matrix of the leptons or the quarks, or whether these are in
the up or down flavour states.  If one likes, this is because 
in the confinement picture of 't Hooft \cite{tHooft} and others 
\cite{Bankovici}, quarks and leptons are bound states via 
$su(2)$ confinement of the fundamental fermion fields $\psi$ 
(which are what carry the up-down flavour and distinguish between 
leptons and quarks) with the weak framon (from which they acquire
their dependence 
on the vector $\balpha$).  This means therefore that the fermion 
matrix (\ref{mfact}) will remain factorized and universal 
even as it rotates with scale, a property that is required in 
\cite{r2m2} to give the mass hierarchy and mixing results we 
seek.  

In addition, these equations are seen to possess a number of 
intriguing features, significant both for theory and for future
phenomenology, which are believed to be generic and to remain 
valid in a more general treatment, and which will now be examined 
in the sections which immediately follow.

\section{The Fixed Points}

The first item of interest is that the equation has a number 
of fixed points which are likely to figure conspicuously in 
the mass spectrum and mixing patterns of quarks and leptons.

There will be a fixed point of the trajectory where $\dot{R}, 
\dot{\theta}$ and $\dot{\phi}$ all vanish.  We note then the 
following:
\begin{itemize}
\item from (\ref{Rdot}), $\dot{R}$ vanishes when $R = 0, -1/2$ 
or $R = 1$ except for $R = 1$ and $\theta = 0$ when $E$ in the 
denominator also vanishes (\ref{E3G});  
\item from (\ref{thetadot}), $\dot{\theta}$ vanishes when $R = 0$ 
or when $\theta = 0, \pi/2$;
\item from (\ref{phidot}), $\dot{\phi}$ vanishes when $\theta = 0$
or when $\phi = 0, \pi/2$;
\end{itemize}
where we have restricted our interest to the first octant of the
unit sphere, the other octants being mere repetitions.  Hence we
conclude that there are fixed points of the trajectory:
\begin{itemize}
\item at $R= 0$ for any values of $\theta$ and $\phi$ ($F0$);  
\item at $R = -1/2$ and $\theta = 0$ for any value of $\phi$ ($F1$); 
\item at $R = 1$, $\theta = \pi/2$, and $\phi = \pi/2$ ($F2$), or $\phi = 0$ ($F3$).
\end{itemize}

If we linearize around the fixed points, taking deviations $\delta_1, \delta_2, \delta_3$ in the $R, \theta, \phi$ directions respectively (at $F3$, e.g., $R=1-\delta_1, \theta=\frac{\pi}{2} -\delta_2, \phi=\delta_3$), we obtain

\begin{eqnarray}
F0: \hspace{1cm}& \dot{\delta}_1& =\frac{\rho_S^2}{4 \pi^2} \delta_1, \ \dot{\theta}=0, \ \dot{\phi} =0\\
F1: \hspace{1cm}& \dot{\delta}_1& = - \frac{5 \rho_S^2}{16 \pi^2}
   \delta_1, \
    \dot{\delta}_2 = - \frac{7 \rho_S^2}{64 \pi^2} \delta_2, \
   \dot{\phi} =0\\
F2: \hspace{1cm}& 
   \dot{\delta}_1& = - \frac{9 \rho_S^2}{32 \pi^2} \delta_1,
 \ \dot{\delta}_2 = - \frac{3 \rho_S^2}{32 \pi^2} \delta_2, \ \ \dot{\delta}_3 
= -\frac{3 \rho_S^2}{32 \pi^2} \delta_3  \\
F3: \hspace{1cm}& 
\dot{\delta}_1& = - \frac{9 \rho_S^2}{32 \pi^2} \delta_1,
 \ \dot{\delta}_2 = - \frac{3 \rho_S^2}{32 \pi^2} \delta_2, \ \ 
 \dot{\delta}_3 = \frac{3 \rho_S^2}{32 \pi^2} \delta_3 
\end{eqnarray}

We can also treat this as an autonomous system of ODEs.  If we evaluate the matrix
\begin{equation}
\left( \begin{array}{ccc}
\partial_R \dot{R}&\partial_\theta \dot{R}&\partial_\phi \dot{R}
\\
\partial_R \dot{\theta}&\partial_\theta \dot{\theta}&\partial_\phi \dot{\theta}
\\
\partial_R \dot{\phi}&\partial_\theta \dot{\phi}&\partial_\phi \dot{\phi}
 \end{array} \right)
\end{equation}
at the fixed point, e.g., $F2$ we get 
\begin{equation}
 \frac{\rho_S^2}{32\pi^2} \left( \begin{array}{ccc}
-9 & 0 & 0 \\
0 & -3 &0\\
0&0 & -3 \end{array} \right)
\end{equation}
and, since the eigenvalues are all negative, it is a stable node.  

From these considerations, one easily concludes that $F0$ is an 
unstable fixed point, $F1$ and $F2$ are stable, and $F3$ is
marginal.  This means that our trajectory will start off at high
scale $\mu = \infty$ near either $F1$ or $F2$, glance off $F3$ if 
it ever gets near that point, and finish eventually at $F0$ at
$\mu = 0$.  For this reason we shall refer to $F1$ and $F2$ as high 
energy fixed points and to $F0$ as the low energy fixed point.  
Notice that the two high energy fixed points $F1$ and $F2$ correspond
to $R$ of different signs.  Recalling the definition of $R$ in
(\ref{R}) above, we see that different signs for $R$ means also
different signs for $\nu_2$ ($\kappa_S$ being by choice positive),
where $\nu_2$ is the coefficient of that term in the framon potential
$V[\Phi]$ of (\ref{VPhi}) linking the strong and weak sectors, which is
responsible for distorting the framons at vacuum from orthonormality.  
A positive sign for $\nu_2$ means attraction, pulling the framon 
axes closer together, and a negative sign means repulsion pushing 
the framon axes further apart.  The two signs, and hence the two 
fixed points, correspond to two different cases, only one of which 
need be considered depending on the choice to fit the physical 
conditions, which we shall have to leave later for phenomenology to 
decide.

\begin{table}[h]
\center
\begin{tabular}{c|c|c|c|c|c|l}
{} & $R$ & $\theta$ & $\phi$ & $\balpha^\dagger$ & $\mu$  & stability
   \\ \hline
$F0$ & $0$ & any &  any& traj. dep. & $0$ & 2 flat directions\\
$F1$ & $-1/2$ & $0$ &  any  & (1,0,0)& $\infty$ & 1 stable + 1 flat\\ 
$F2$ & $+1$ & $\pi/2$ & $\pi/2$ & (0,1,0)& $\infty$ & stable\\ 
$F3$ & $+1$ & $\pi/2$ & $0$ & (0,0,1)& $\infty$ & 2 stable + 1 unstable\\ 
\end{tabular} 
\caption{RGE fixed points} 
\label{fixedpts}
\end{table}

Our conclusions above for the fixed points are summarized in Table
\ref{fixedpts}.  That two particular values of $\phi$, namely $0$
and $\pi/2$, as listed, should be picked out as fixed points may
seem surprising given the original symmetry of the problem under
rotation about the $\theta = 0$ axis.  This will not be the case
when we recall the fact that there still remains in the problem an
arbitrary rotation called $R_1'$ in (\ref{Ainv3Gdp}) above by which the
trajectory can be rigidly rotated about the $\theta = 0$ axis.  By
means of this rotation, the two fixed points $F2$ and $F3$ can
be placed at any values of $\phi$, so long as these stay $\pi/2$
apart.

Although the fixed points listed in the Table \ref{fixedpts} were
deduced above from the equations (\ref{Rdot})---(\ref{phidot}),
which themselves were derived with only one (strong) Higgs loop,
we notice that they all correspond to very special locations of
the rotation trajectory, either when the framons are orthogonal
($R = 1, - 1/2$) though not of equal lengths, or else when the
framons are actually orthonormal ($R = 0$), as
can be seen from (\ref{Phivac}) and (\ref{Phivac0}) above.  And
this conclusion is a consequence directly of the form of the
framon potential (\ref{VPhi}), not of the subsequent one-loop
approximation made in deriving the rotation equations.  One
believes therefore that the presence of these fixed points is
generic and much more general than its derivation given here.  
 
That there are fixed points on the trajectory for ${\balpha}$ is 
of crucial importance for the success of the FSM, and indeed of 
any rotation model, in explaining  mass hierarchy and mixing for 
the following reason.  The idea all along is that both mixing 
and the lower generation masses come from rotation.  So, when 
$\balpha$ approaches a fixed point, rotation will slow down 
progressively and give smaller and smaller effects.  This is 
easiest to visualize at the high scale end, where the existence 
of a fixed point would predict that the mass leakage to lower 
generations will become progressively smaller the higher the 
mass scale, hence $m_c/m_t < m_s/m_b < m_\mu/ m_\tau$, as is 
experimentally observed.  Indeed, the proximity of the heavier 
quark states to the high energy fixed point is such that the 
rotation angles involved are small enough for some well-known 
differential formulae \cite{Darboux} of space curves to apply 
to the rotation trajectory, leading immediately to most of the 
salient features in the CKM matrix \cite{cornerel} which are 
observed in experiment \cite{databook}.  These effects are so 
pronounced that even a glance at the data, when interpreted in 
terms of rotation, would already suggest the existence of such 
an asymptote \cite{cevidsm,r2m2}.  Furthermore, the existence 
of a rotational fixed point at $\mu = \infty$ would imply that 
mixing angles are in general smaller for quarks, these being 
heavier, and therefore nearer to the fixed point, than for 
leptons, which is again as observed in the experimental CKM and 
PMNS matrices.  The other fixed points on the trajectory listed
in Table 1 are of perhaps no less significance but their physical 
implications are not yet entirely clear to us.

\section{The Scale-Dependent Metric}

The equations (\ref{Rdot}), (\ref{thetadot}) and (\ref{phidot})
governing the rotation of the vector ${\balpha}$ are derived
from the manner the vacuum changes under a change of scale, where 
the vacuum at any scale in turn specifies a set of values for the 
strong framons $\Phi$.  The latter, having themselves been given
the geometric significance of frame vectors to begin with, will 
then specify a metric.  We conclude therefore that the metric 
too will depend on scale.  The purpose of this section is to 
clarify this dependence and some of its physical implications.

Recall first that we have started with a theory invariant under
$su(3) \times \widetilde{su}(3)$, but, the vacuum being degenerate,
the choice of a particular vacuum breaks this symmetry.  Our
contention, as argued in \cite{efgt}, was that colour is confining
so that the local $su(3)$ symmetry should still be exact.  What 
is broken by the choice of vacuum is thus only the global symmetry
$\widetilde{su}(3)$.  The breaking of this $\widetilde{su}(3)$ is 
reflected in the fact 
that the vacuum values of the strong framon field
$\Phi$, as given in (\ref{Phivac}) above, are distorted from 
orthonormality, or that the metric is no longer flat.  We wish now 
to find out explicitly what form this departure from flatness of 
the metric in $\widetilde{su}(3)$ will take.

Let us start with the reference vacuum in which the value of $\Phi$
takes the particularly simple form in (\ref{Phivac0}). This gives
the metric in $\widetilde{su}(3)$ as usual as
\begin{equation}
g^{\tilde{a} \tilde{b}} = \sum_{a} (\phi_a^{\tilde{a}})^*
                                    \phi_a^{\tilde{b}},
\label{glower}
\end{equation}
or in matrix form as
\begin{equation}
\tilde{G}^0 = (\phivac^0)^\dagger \phivac^0
            = \frac{\zeta_S^2}{3} \left( \begin{array}{ccc} 
                                      1+2R & 0 & 0 \\
                                      0 & 1-R & 0 \\
                                      0 & 0 & 1-R \end{array} \right),
\label{G0up}
\end{equation}
which we see is still diagonal but no longer flat.

Next, we recall from (\ref{Phivac}) that the vacuum value $\phivac$  
for the general vacuum $\Phi$ is obtainable 
from that of the reference vacuum $\phivac^0$  by an 
$\widetilde{su}(3)$ transformation $A^{-1}$ from the right, 
so that for the general vacuum,
\begin{equation}
\tilde{G} = A \tilde{G}^0 A^{-1},
\label{Gup}
\end{equation}
which is now not even diagonal.  

As the scale $\mu$ changes, both the matrix $A$ and the quantity 
$R$ in $\tilde{G}^0$ will change, and so will the metric.  It 
thus follows that in evaluating any metric-dependent quantities, 
such as lengths and inner (dot) or outer (cross) products
of vectors in generation space, a metric at the appropriate scale
will have to be adopted.  And as noted already in 
(\ref{Utriad}) and (\ref{hiermass}) in \S2, such quantities are 
required in the 
rotation scenario for calculating the masses, state vectors and 
mixing angles of the various quark and lepton states.

To evaluate the lengths or the products of two vectors at the
same scale, one takes the metric 
at that scale.  As our notation goes, where vectors carry
generation indices as superscripts, it is the inverse of the 
matrix (\ref{Gup}) above that we need to use to evaluate the 
inner products of vectors, thus
\begin{equation}
G = 3 \zeta_S^{-2}(\mu) A(\mu) \left( \begin{array}{ccc} 
              (1 + 2R(\mu))^{-1} & 0 & 0 \\
              0 & (1 - R(\mu))^{-1} & 0 \\
              0 & 0 & (1 - R(\mu))^{-1} 
           \end{array} \right) A^{-1}(\mu).
\label{Gdown}
\end{equation}
However, to evaluate the 
product between two vectors defined at two different scales, we 
need further clarification.  For example, according to \S2 and 
\cite{r2m2}, the CKM matrix element $V_{tb}$ is the inner product 
between the state vector ${\bf t} = {\balpha}(\mu = m_t)$ of $t$ 
and the state vector ${\bf b} = {\balpha}(\mu = m_b)$ of $b$.  
Since now the metric depends on $\mu$, one has to specify 
which metric is to be used to evaluate the inner product.  This 
situation is, however, familiar in gravity where the metric also 
varies from point to point in space-time and one has to specify 
what is meant by the same (or parallel) vector at different
space-time points.  It is for this that the geometrical concept 
of parallel transport is introduced.  In gravity, it is the 
Christoffel symbols usually denoted $\Gamma^c_{ab}$ which tell us 
what are parallel vectors at two neighbouring points, which, once 
known, can be repeated to specify what are parallel vectors along 
any curve even at a finite distance apart.  To calculate then the 
inner product between two vectors defined at a finite distance 
from each other along a curve, one can parallelly transport both 
vectors along the curve to the same point and take their inner 
product with respect to the local metric valid there.  This inner 
product is symmetric and invariant under parallel transport, as 
it should be.  

In view of this, the answer to our specific question above is then 
clear.  To calculate $V_{tb}$, we take ${\bf t}$ and parallelly
transport it along the rotation trajectory from $\mu = m_t$ to 
$\mu = m_b$, then take its inner product with ${\bf b}$ using 
the metric at $\mu = m_b$.  Indeed, since the inner product is 
symmetric and invariant under parallel transport, one can equally
well do the reverse, namely parallelly transport ${\bf b}$ from
$\mu = m_b$ to $\mu = m_t$ and take its inner product there with 
${\bf t}$ using the metric at $\mu = m_t$, or else parallelly
transport
both vectors ${\bf t}$ and ${\bf b}$ to an arbitrary common
scale and evaluate their inner product using the metric valid
there.  The answer will be the same.

The only question left is how exactly to effect parallel transport 
in our system, or in other words, what are our equivalents of the 
Christoffel symbols in gravity.  A theorem in metric geometry says 
that if a metric is covariantly constant (inner products invariant 
under parallel transport) and torsion free (the metric $g_{ab}$ 
is symmetric), then the Christoffel symbols are given in terms of 
the metric by the formula
\begin{equation}
\Gamma^c_{ab} = \textstyle{\frac{1}{2}} g^{cd} (\partial_a g_{db} + 
\partial_b g_{ad} - \partial_d g_{ab}),
\label{Christoffel}
\end{equation}
familiar in gravity.  This can be applied to our system here,
treating $\widetilde{su}(3)$ as ``space'' and the scale $\mu$ as the
``time-coordinate'', to deduce the corresponding Christoffel symbol 
and hence parallel transport, as is done in Appendix B.  However,
our system being so simple, it is not hard to guess the answer
directly, as we actually did first, without going through this
calculation.  To parallelly transport a vector from the point $\mu$ 
to another point $\mu'$ along a trajectory parametrized by $\mu$,
one needs just to multiply the vector by the following matrix 
\begin{eqnarray}
\Pi(\mu \rightarrow \mu') & = & A(\mu') \left( \begin{array}{ccc} 
                      \zeta_S' P' & 0 & 0 \\
                      0 & \zeta_S' Q' & 0 \\
                      0 & 0 & \zeta_S' Q' \end{array} \right)
                      A^{-1}(\mu') \nonumber \\
                      & & \times A(\mu) \left( \begin{array}{ccc}
                      (\zeta_S P)^{-1} & 0 & 0 \\
                      0 & (\zeta_S Q)^{-1} & 0 \\
                      0 & 0 & (\zeta_S Q)^{-1} \end{array} \right)
                      A^{-1}(\mu),
\label{paratrans}
\end{eqnarray}
where we have introduced the shorthand notation
\begin {equation}
P = \sqrt{\frac{1 + 2R}{3}},\ \ \  Q= \sqrt{\frac{1 - R}{3}},
\label{PQ}
\end{equation}
two quantities which will occur frequently in what follows.
This is easily seen to preserve inner products between vectors 
which is, after all, the essential element of the above cited 
theorem.  Then with parallel transport as given by $\Pi$ in 
(\ref{paratrans}) above, it is easy now to evaluate the inner 
product between two vectors even defined at two different scales.  

The cross product between vectors, say ${\bf a}$ and ${\bf b}$, 
defined at different scales can most easily  be evaluated as 
follows.  We first parallelly transport by $\Pi$ in (\ref{paratrans}) 
both vectors ${\bf a}$ and ${\bf b}$ from the scales where they are 
defined to the scale corresponding to $R =0$ where we see from 
(\ref{Gdown}) that the metric is flat.  We then take the cross 
product of the two transported vectors there with respect to the 
flat metric, thus
\begin{equation}
c^i = \epsilon^{ijk} a^j b^k.
\end{equation}
The cross product {\bf c} can then be parallelly transported to 
any desired scale by (\ref{paratrans}) which, as already noted, 
will preserve both its length and orthogonality with ${\bf a}$ and 
${\bf b}$ parallelly transported to the same scale with respect 
to the metric appropriate for that scale.

Having now worked out how the norms and the products (both dot and 
cross) of vectors in generation space are to be taken, the latter 
even between vectors defined at different scales, one can proceed
now to evaluate the masses and mixing matrices of both leptons and 
quarks according to the rules summarized in \S2 with the effects 
of the metric folded in.  It would appear at first sight that this
might affect much the previous conclusions, e.g., in \cite{r2m2} 
deduced with the flat metric, but this turns out surprisingly not
to be the case.  

To be explicit, let us introduce at every scale $\mu$ as our ``local'' 
reference frame the ``Darboux triad'' \cite{Darboux} consisting of,
first, the vector $\balpha(\mu)$, secondly the tangent vector to
the trajectory $\btau(\mu)$ at that scale, and thirdly the normal 
$\bnu(\mu)$ to both the above, all three being normalized and
mutually orthogonal with respect to the original flat metric.  We
take also the matrix $A(\mu)$ explicitly to be that matrix which 
takes the reference vectors at the reference vacuum to the Darboux
triad at $\mu$, thus
\begin{equation}
\balpha(\mu) = A(\mu) \left( \begin{array}{c} 1 \\ 0 \\ 0 
               \end{array} \right), \ \ 
\btau(\mu) =  A(\mu) \left( \begin{array}{c} 0 \\ 1 \\ 0 
               \end{array} \right), \ \
\bnu(\mu) =  A(\mu) \left( \begin{array}{c} 0 \\ 0 \\ 1 
               \end{array} \right).
\label{Aexplicit}
\end{equation}
With respect to the ``local'' metric at $\mu$, namely (\ref{Gdown}),
the Darboux triad is no longer normalized
\begin{equation}
\langle \balpha|\balpha \rangle = \zeta_S^{-2} P^{-2}, \ \ 
\langle \btau|\btau \rangle = \zeta_S^{-2}Q^{-2}, \ \
\langle \bnu|\bnu \rangle = \zeta_S^{-2} Q^{-2},
\label{normDarboux}
\end{equation}
but remains orthogonal by virtue of the special form of the metric  
(\ref{Gdown}).

The state vector ${\bf t}$ of $t$ is defined at $\mu = m_t$ to be,
as before, the vector parallel to $\balpha$, but has now to be
normalized with respect to the local metric at $\mu$, hence
\begin{equation}
{\bf t} = \zeta_{St} P_t \balpha(\mu = m_t)
    = \zeta_{St} P_t A(\mu = m_t) \left( \begin{array}{c}
      1 \\ 0 \\ 0 \end{array} \right),
\label{vtm}
\end{equation}
where a subscript, $t$ say, on scalar quantities such as $\zeta_S$ 
and $P$ denotes their values evaluated at $\mu = m_t$, but vector
quantities which are subject to parallel transport have the value 
of $\mu$ at which they are evaluated explicitly stated.  The state 
vectors ${\bf c}$ of $c$ and ${\bf u}$ of $u$ remain both orthogonal 
to ${\bf t}$ and to each other by (\ref{Gdown}), so that 
\begin{equation}
{\bf c} \propto \Omega_U \btau, \ \ {\bf u} \propto \Omega_U \bnu,
\label{vcvuprop}
\end{equation}
where $\Omega_U$ is a rotation about the ${\bf t}$ vector:
\begin{eqnarray}
\Omega_U & = & A(\mu = m_t) \left( \begin{array}{ccc}
   1 & 0 & 0 \\ 0 & \cos \omega_U & - \sin \omega_U \\ 
   0 & \sin \omega_U & \cos \omega_U \end{array} \right)
   A^{-1}(\mu = m_t) \nonumber \\ 
         & = & A(\mu = m_t) \,\Omega_U^0\, A^{-1}(\mu = m_t).
\label{OmegaU}
\end{eqnarray}
Normalizing then the vectors in (\ref{vcvuprop}) with respect to
the local metric at $\mu = m_t$, we have
\begin{eqnarray}
{\bf c} & = & \zeta_{St} Q_t A(\mu = m_t) \Omega_U^0 \left( 
   \begin{array}{c} 0 \\ 1 \\ 0 \end{array} \right), \nonumber \\ 
{\bf u} & = & \zeta_{St} Q_t A(\mu = m_t) \Omega_U^0 \left( 
   \begin{array}{c} 0 \\ 0 \\ 1 \end{array} \right).
\label{vcvum}
\end{eqnarray}

Applying the same arguments to the $D$ quarks as we did to the 
$U$ quarks above, and introducing the corresponding rotation matrix
$\Omega_D^0$,  we obtain
\begin{eqnarray}
{\bf b} & = & \zeta_{Sb} P_b A(\mu = m_b) \left( 
   \begin{array}{c} 1 \\ 0 \\ 0 \end{array} \right), \nonumber \\
{\bf s} & = & \zeta_{Sb} Q_b A(\mu = m_b) \Omega_D^0 \left( 
   \begin{array}{c} 0 \\ 1 \\ 0 \end{array} \right), \nonumber \\ 
{\bf d} & = & \zeta_{Sb} Q_b A(\mu = m_b) \Omega_D^0 \left( 
   \begin{array}{c} 0 \\ 0 \\ 1 \end{array} \right).
\label{vDm}
\end{eqnarray}

The mixing elements in the CKM matrix are given as before as the 
inner products between the state vectors of the $U$ and $D$ quarks,
only now with the difference that, the $U$ and $D$ state vectors
being defined at different scales, they have first to be parallelly
transported to the same scale, and their
inner products have to be evaluated with the metric (\ref{Gdown})
appropriate for that scale.  For instance, as discussed above, 
to evaluate the CKM 
element $V_{tb}$, we can parallelly transport the vector ${\bf t}$
in (\ref{vtm}) from the scale $\mu = m_t$ to $\mu = m_b$, thus
\begin{equation}
{\bf t}(\mu \rightarrow m_b) = \Pi(\mu = m_t \rightarrow \mu = m_b)
                   {\bf t},
\label{vtatmb}
\end{equation}
using the parallel transport
operator given in (\ref{paratrans}), and then take its inner
product with the vector ${\bf b}$ defined at $\mu = m_b$ with respect 
to the metric (\ref{Gdown}) at the scale $\mu = m_b$.  Hence
\begin{equation}
V_{tb} = {\bf t}^\dagger(\mu \rightarrow m_b) A(\mu = m_b) \left( 
        \begin{array}{ccc} \zeta_{Sb}^{-2} P_b^{-2} & 0 & 0 \\
                           0 & \zeta_{Sb}^{-2} Q_b^{-2} & 0 \\
                           0 & 0 & \zeta_{Sb}^{-2} Q_b^{-2}
        \end{array} \right) A^{-1}(\mu = m_b) {\bf b}.
\label{Vtba}
\end{equation}
Substituting the expressions obtained before, one then easily
obtains that
\begin{equation}
V_{tb} = (1, 0, 0) A^{-1}(\mu = m_t) A(\mu = m_b)
         \left( \begin{array}{c} 1 \\ 0 \\ 0 \end{array} \right)
       = {\bf t} \cdot{\bf b},
\label{Vtbb}
\end{equation}
namely, exactly the same answer as was given before in (\ref{VCKM})
without incorporating the scale-dependent non-flat metric.

This conclusion that the CKM matrix element remains formally the 
same in terms of the state vectors with or without incorporating 
the scale-dependent metric holds not just between the $t$ and $b$ 
states as above demonstrated, but between any pair of $U$ and $D$ 
states, as can readily be checked explicitly.  The reason for such
a simple answer is that the ``local'' metric at $\mu$ is diagonal in
these state vectors, so that whether in parallel transport or in
forming the inner product, the vectors just get simply multiplied 
by some factors of $\zeta_S P$ or $\zeta_S Q$, and these 
all eventually cancel out.  

However, this result by itself does not yet mean that the actual 
values of the CKM matrix will remain the same with or without the
scale-dependent metric, for it is still to be verified how the
state vectors themselves will be affected by the introduction of
the scale-dependent metric.  To see this, let us work out as an
example explicitly the $U$ states.  We recall from (\ref{mfact})
that the mass matrix of the $U$ quarks is scale-dependent, so 
that at $\mu = m_c$, relevant for the evaluation of the state
vector of $c$, the mass matrix reads as
\begin{equation}
m(\mu = m_c) = m_U\,
                \balpha(\mu = m_c) \balpha^\dagger(\mu = m_c).
\label{matmc}
\end{equation}
According to the analysis in \S2, the physical mass $m_c$ for the 
$c$ quark is given by the diagonal element of $m(\mu = m_c)$ 
taken between the state vector of the $c$ quark.  But now, for 
the case of the scale-dependent metric, the vector ${\bf c}$ has 
first to be parallelly transported from $\mu = m_t$ where it was 
originally defined in (\ref{vcvum}) to $\mu = m_c$, and its matrix
element of $m(\mu = m_c)$ has to be evaluated with respect to the 
metric (\ref{Gdown}) taken again at $\mu = m_c$.  These by now 
familiar operations then yield
\begin{equation}
m_c = m_U\, \zeta_{Sc}^{-2}\, P_c^{-2}\,
         |\balpha(\mu = m_c)\cdot{\bf c}(\mu \rightarrow m_c) |^2,
\label{mcma}
\end{equation}
an answer differing from that obtained before in (\ref{hiermass})
in \S2 without the scale-dependent metric merely by a factor 
$\zeta_{Sc}^{-2} P_c^{-2}$.  Besides, a repetition of the argument
for the masses of the $t$ and $u$ quarks gives the same factor,
only now taken at respectively the scales $m_t$ and $m_u$.

In other words, as far as the calculation of the quark masses is 
concerned, what the scale-dependent metric has done is to multiply 
the coefficient $m_T(\mu)$ by the factor $\zeta_S^{-2}(\mu)
P^{-2}(\mu)$.  And since in the rotation scheme, the state vectors
of the lower generations, such as ${\bf c}$ and ${\bf u}$, also 
depend on the mass calculation, then even the CKM matrix would 
be affected by the introduction of the scale-dependent metric.  

At least, that would be the case in theory if we know what the 
value of $m_T$ is and how it varies with $\mu$.  However, at the
phenomenological level at which present fits to experiment are
performed, as in \cite{r2m2}, the coefficient $m_T$ is treated as 
empirical to be fitted to data.  Then, it would not matter whether
it was $m_T$ or $m_T \zeta^{-2}(\mu) P^{-2}(\mu)$ that is to 
be fitted empirically, and the two cases, with or without the
scale-dependent metric, would yield the same answer; i.e., the
phenomenology reviewed in \cite{r2m2} would still apply now in
the present case in FSM with a scale-dependent metric.  Thus, for 
example, in \cite{r2m2} fairly good fits were obtained assuming $m_T$
to be approximately independent of $\mu$ when the metric was taken
to be flat and $\mu$-independent.  Then the same fit would be 
obtained here with the $\mu$-dependent non-flat metric assuming
instead that
$m_T \zeta^{-2}(\mu) P^{-2}(\mu)$ is approximately $\mu$-independent.  
Notice, however, that this statement has been shown to be valid 
only when we concern ourselves just with the mass hierarchy and 
the mixing pattern.  One would hope that in probing further into 
 physical phenomena beyond the above limited domain, then the 
effect of the scale-dependent metric may make itself manifest, 
but of this we have as yet found no clear example.

We end this section by noting that although we have performed
the analysis on the scale-dependent metric by starting with the
rotation equation (\ref{Rdot})---(\ref{phidot}), very little of 
the result depend in fact on them.  The form of the metric
(\ref{Gdown}) and of the parallel transport (\ref{paratrans}) in
terms of the quantity $R$ is a consequence merely of the framon
potential (\ref{VPhi}), which is in turn the consequence of 
the double invariance under $su(3) \times su(2) \times u(1)$ and 
$\widetilde{su}(3) \times \widetilde{su}(2) \times \tilde{u}(1)$ 
plus renormalizability.  And from these premises already all the 
results discussed would follow.  Only the details of how $R$ 
actually varies with $\mu$ would depend on the rotation equations, 
such as those above.  We stress therefore that the result deduced 
above for the scale-dependent metric is generic for the FSM scheme 
and not subject to the limitations of the approximations used to
derive the equations (\ref{Rdot})---(\ref{phidot}).

\section{The KM CP-violating Phase}

One special feature of the rotation equation derived in \S6 for 
the vector ${\balpha}$ is that the phases of the elements remain 
unchanged with changing $\mu$.  Since the state vectors of the 
various fermion states are all themselves derived eventually
from ${\balpha}$, though each at some specific value of $\mu$,
the above observation would mean that their elements would carry 
the same phases also.  Hence, in taking the inner products between
these state vectors to calculate the mixing matrices according
to (\ref{VCKM}) for quarks and a similar expression for leptons, 
these phases will all cancel and one will obtain real values for 
all entries.  In other words, these mixing matrices will have no 
Kobayashi-Maskawa phase \cite{KM} and be CP-conserving.  
If this were to be 
the final answer---and for some time we thought it was---then it 
would be disappointing, for the possibility of having such a 
CP-violating phase is one of the most intriguing properties of 
the 3-generation mixing matrix.

At first sight, it might appear that the above result is just an
accident of the particular manner the equation was derived, namely
from the insertion of a single strong Higgs-loop into the fermion
self-energy as specified in section 5.  On reflection, it is soon
realized that this is not the case.  We recall that the idea all
along is that rotation is driven by renormalization effects in the
strong sector and only gets transmitted to ${\balpha}$ in the
weak sector via the linkage term $V_{WS}$ in the framon potential
$V[\Phi]$  in (\ref{VPhi}).  And it is this rotation which gives 
rise to the CKM and PMNS mixing matrices.  Thus, if these matrices 
were to develop a Kobayashi-Maskawa phase and hence CP-violations, 
it would mean that strong interactions where the effect originates, 
though CP-conserving themselves, are capable somehow of generating 
CP-violating effects via rotation in the weak sector.  This does 
not seem reasonable.  It would appear that for such a mechanism 
to give CP-violating phases in the mixing matrices, one will have 
to start with a framework where the strong interactions themselves 
are CP-violating.

Surprisingly, this last conclusion is not as hopeless as it might
seem.  We have to recall first that strong interaction as embodied 
in QCD is {\it a priori} not CP-conserving since gauge and Lorentz 
invariance admit in QCD in principle a CP-violating term of the 
form
\begin{equation}
{\cal L}_{\theta} = - \frac{\theta}{64 \pi^2} 
   \epsilon^{\mu \nu \rho \sigma} F_{\mu \nu} F_{\rho \sigma}
\label{Ltheta}
\end{equation}
of topological origin, where $\theta$ can take any arbitrary value
\cite{Weinbergbook}.
Simply because of the absence of any strong CP-violations observed
in experiment \cite{edm}, it is customary to declare by fiat that $\theta$ in
(\ref{Ltheta}) above is zero, or to explain why it has to be less than
about $3 \times 10^{-10}$.
The need to do so is in fact 
known as the strong CP problem, a classic problem that has been 
with us for more than 40 years \cite{Weinbergbook}.  
It would be much more satisfying 
theoretically if one could start instead with the general action 
including a theta-angle term, with the coefficient $\theta$ not
vanishingly small,
that the original invariance principles 
allow and find some theoretical reason why it would not necessarily 
lead to strong CP-violations in contradiction to experiment.  This 
is what is meant in common usage by a solution to the strong 
CP-problem.

Now a very attractive feature of the rank-one rotating mass matrix 
(R2M2) mechanism, for which the present FSM is an example, is that, 
in addition to offering an explanation for the distinctive 
fermion mass and mixing patterns observed in experiment 
as outlined in the introduction, it offers as a by-product
also a neat solution to the strong CP problem, transforming
the unwanted theta-angle into a CP-violating phase in the 
CKM matrix where it is actually wanted.  How this comes about is 
as follows.  

It has long been known that if one were to make a chiral 
transformation on a fermionic variable, thus
\begin{equation}
\psi \rightarrow \exp(i \alpha \gamma_5) \psi,
\label{chiraltrans}
\end{equation}
then the Feynman integral will acquire from the Jacobian 
of the transformation a factor of the same form as the
theta-angle term (\ref{Ltheta}) above, only with $\theta$
there replaced by $2 \alpha$.  Since physics should not be
changed by a change in integration variables it follows 
that any theta-angle term at first present in the action
can thus be eliminated by a judicious chiral transformation on 
the quark fields.  The trouble, however, is that the chiral
transformation will affect also other terms in the action
depending on $\psi$, in particular the quark mass term, 
which in general will go complex, thus
\begin{equation}
m \bar{\psi} \psi \rightarrow m \exp(2 i \alpha) \bar{\psi}
   \half (1 + \gamma_5) \psi + m \exp(-2 i \alpha)
   \bar{\psi} \half (1 - \gamma_5) \psi.
\label{mchitrans}
\end{equation}
This would normally mean CP-violations again, unless $m$ 
happens to be zero.  But, as far as we understand
at present, experiment does not seem to want any quarks to
have zero mass. 

It is at this point that R2M2 starts to make a difference.
We recall that the fermion mass matrix there is of the form
(\ref{mfact}) which has 2 zero eigenvalues at every $\mu$.
A chiral transformation can thus be performed on either of
these eigenstates without making the mass term complex, as
per (\ref{mchitrans}).  In other words, at any $\mu$, any 
theta-angle term in the action can be eliminated by a chiral 
transformation without making the mass term complex.  Yet,
as outlined in \S2 and explained in more detail in, for 
example, \cite{r2m2}, this does not require any of the
quarks to have physical zero mass since they can all acquire masses
by the ``leakage'' mechanism because of rotation, avoiding 
thus any conflict as yet with experiment.

But this is not all.  One still has to check whether the
chiral transformation performed to eliminate $\theta$ will
affect other terms in the action and lead to CP-violations
elsewhere.  The interesting thing is that it does, but only
in just the right place where it is needed.  As analysed in
\cite{atof2cps}, to keep the mass matrix hermitian at all 
$\mu$, the chiral transformation for eliminating $\theta$ 
has to be performed on the state orthogonal to both 
the rotating vector ${\balpha}$ and the tangent to the 
rotation trajectory at every $\mu$.  But this normal 
direction ${\bnu}(\mu)$ is itself also $\mu$-dependent 
because of rotation, so that the CKM matrix (\ref{VCKM}) 
which involves vectors defined at different $\mu$'s, will
 acquire thereby new phases from the chiral transformations.  
In other words, the R2M2 mechanism allows the elimination 
of the theta-angle term, i.e.\ a solution of the strong CP 
problem, without making the mass term complex, but only at 
the cost of introducing a CP-violating phase in the CKM matrix, 
even with ${\balpha}$ real to begin with.  But this is, of 
course, a price one is most willing to pay, for this 
phase was exactly what one was looking for above at the 
beginning of the section.  

What is perhaps most gratifying for the rotation scheme, with
the above mechanism for generating the Kobayash-Maskawa phase 
in the CKM matrix, is that it even yields CP-violations of the 
correct order of magnitude.  It was shown in \cite{atof2cps,r2m2} that starting with a theta-angle of order unity in the 
strong sector, the rotation scheme will automatically end up 
with a Jarlskog invariant \cite{Jarlskog} of order $10^{-5}$ 
as is observed in experiment \cite{databook} provided that
the rotation is adjusted to yield roughly the correct value
for say $m_c/m_t$ by leakage as per (\ref{mc2}). 

The analysis of \cite{r2m2,atof2cps}, of which the above is 
a brief paraphrase, applies in general terms to the situation 
here in FSM, but in detail has to be modified.  The reason
is that the analysis there was based on the assumption that
in generation space the metric is flat.  Here, in the FSM, as
detailed in \S8 the metric is not flat.  Hence, any quantity, 
such as lengths and products of vectors used in the analysis 
will have to be recalculated here in terms of the FSM metric.  
However, rather than repeating the detailed analysis given in 
\cite{r2m2}, inserting the FSM metric wherever appropriate, 
it would be sufficient here and quicker to take a new tack 
to indicate how the effect can be calculated.

We shall do so first for the situation when the metric is
flat.  We recall then that the Darboux triad, set up above in 
(\ref{Aexplicit}) and consisting of the 3 vectors ${\balpha}$, 
$\btau$ and $\bnu$, forms an orthonormal basis in generation 
space at every point $\mu$ of the rotation trajectory.  At 
every $\mu$, according to the preceding analysis, a chiral 
transformation is to be performed to eliminate the theta-angle 
term on the state in the direction of ${\bnu}$ so as to keep 
both $m(\mu)$ and $m(\mu+\delta \mu)$ hermitian \cite{atof2cps}.  
Since we are now concentrating on the CKM mixing matrix, where 
only left-handed fields occur, we can replace the chiral
transformation (\ref{chiraltrans}) by just the phase factor 
$\exp (-i \theta/2)$.

Writing out then the state vectors, say, of the $U$-quarks in 
terms of the Darboux triad as basis at $\mu = m_t$, we have
\begin{eqnarray}
\tilde{{\bf t}} & = & \balpha(\mu = m_t), \nonumber \\
\tilde{{\bf c}} & = & \cos \omega_U \btau(\mu = m_t) + \sin \omega_U
              \bnu(\mu = m_t) e^{-i \theta/2}, \nonumber \\ 
\tilde{{\bf u}} & = & -\sin \omega_U \btau(\mu = m_t) + \cos \omega_U
              \bnu(\mu = m_t) e^{-i \theta/2},
\label{tcutilde}
\end{eqnarray}
using the notation introduced in (\ref{OmegaU}) above.  Similar 
expressions are obtained for the D-type quarks with $\omega_U$ 
changed to $\omega_D$ and the Darboux triad evaluated instead
at $\mu = m_b$.
 
The CKM mixing matrix can now be expressed as the inner products of 
the chirally rotated quark states (cf. equation (\ref{VCKM}))
\begin{equation}
V_{CKM} = \left( \begin{array}{ccc}
   \tilde{\bf u} \cdot\tilde{\bf d}  &  \tilde{\bf u} \cdot
\tilde{\bf s}  &  \tilde{\bf u} \cdot \tilde{\bf b}  \\
    \tilde{\bf c} \cdot \tilde{\bf d}  &  \tilde{\bf c} \cdot
\tilde{\bf s}  &  \tilde{\bf c} \cdot \tilde{\bf b}  \\
    \tilde{\bf t} \cdot \tilde{\bf d}  &  \tilde{\bf t} \cdot
\tilde{\bf s}  &  \tilde{\bf t} \cdot \tilde{\bf b} 
          \end{array} \right).
\label{CKMtilde}
\end{equation}
And because the direction $\bnu$ in which the chiral phase occurs
varies with scale $\mu$, and the $U$ and $D$ vectors are evaluated
at different scales, i.e., $m_t$ and $m_b$ respectively, the inner
products appearing in (\ref{CKMtilde}) will in general be complex 
therefore leading to a nonvanishing Jarlskog invariant and hence 
CP-violation, as concluded before in \cite{atof2cps,btfit}.

What happens now when the metric is not flat in the present FSM
situation?  With respect to the FSM metric, as already noted, 
none of the vectors in the Darboux triad are now of unit length
although they remain mutually orthogonal.  Nevertheless, one 
can construct a new triad orthonormal with respect to the FSM
metric as follows:
\begin{eqnarray}
{\brho}' & = & {\balpha}/|{\balpha}| \nonumber \\ 
{\bnu}'  & = & \frac{{\balpha} \times {\btau}}
   {|{\balpha} \times {\btau}|} \nonumber \\
{\btau}' & = & {\bnu}' \times {\brho}',
\label{Darbouxp}
\end{eqnarray}
where all lengths and products are to be evaluated in terms of 
the FSM metric.  Again, the 
state vectors of the various quark states, as constructed in the
preceding section with respect to the FSM metric, can be written 
out in terms of the 3 vectors ${\brho}',{\btau}',{\bnu}'$ as in
(\ref{tcutilde}) above except that every vector will now have to 
be primed to indicate that it is defined with respect to 
the non-flat FSM metric.  The same applies to both the $U$ and
the $D$ quarks, from which one concludes that the CKM matrix too
will look the same as in (\ref{CKMtilde}) above, with only the
proviso that all inner products are to be evaluated with the 
FSM metric and that the two vectors involved in the product have 
first to be parallelly transported to a common scale 
before the product is taken.

The amusing thing is that, as noted in the preceding section at
the end, these inner products actually appear the same whether 
evaluated with or without the FSM metric.  This then means that 
not only the conclusion that elimination of the theta-angle will
lead to a Kobayashi-Maskawa CP-violating phase in the CKM matrix,
but also the result that it will lead to a Jarlskog invariant of
the same order of magnitude as observed in experiment for a
theta-angle of order unity,
obtained
before in \cite{atof2cps} with a flat metric, will both still 
be preserved in the present case with the FSM metric.  And this
result is again generic, dependent only on the properties of 
the vacuum, not on the approximations on which the particular
rotation equations (\ref{Rdot})---(\ref{phidot}) were derived.

\section{Summary and Remarks}

Let us first briefly summarize what seems to have been achieved
by formulating the standard model as a framed gauge theory as
has been done in \cite{efgt} and developed in this paper.

By its very nature as frame vectors, framon fields carry in
addition to indices referring to the local gauge symmetries
$su(3)$, $su(2)$, and $u(1)$, also indices referring to the 
global symmetries $\widetilde{su}(3)$, $\widetilde{su}(2)$, and
$\tilde{u}(1)$.  The action for the framed standard model is
to be invariant under both these local and global symmetries.
The occurrence of the 3 global symmetries are welcome since they 
can play the role of fermion generations, up-down flavour, and 
baryons-lepton number respectively \cite{efgt}, while offering 
a geometric interpretation for them which was previously 
unavailable in the usual formulation of the standard model.

The scalar framon fields introduced by (minimal) framing are 
of two types \cite{efgt}, weak and strong.  The weak
framon is of the form $\phi_r^{\tilde{r} \tilde{a}} = \alpha^
{\tilde{a}} \phi_r^{\tilde{r}}$.  It contains a global factor 
$\balpha$, a vector in 3-D generation space, in addition to the
scalar field $\phi_r^{\tilde{r}}$, which is basically the same 
as that which occurs in the standard electroweak theory.  Hence,
both leptons and quarks which are, in the confinement picture 
of 't Hooft \cite{tHooft} and others \cite{Bankovici}, bound 
states of the weak framon with fundamental fermion fields, also
carry the global factor $\balpha$ and acquire thereby the index
$\tilde{a}$ to play the role of the generation index.

The Yukawa couplings constructed with the weak framon will thus
automatically give rise to mass matrices of the factorizable 
(rank-one) form (\ref{mfact}) for both quarks and leptons, with
$\balpha$ as a factor, which is universal, being a property
of the framon, not of the fermion to which it is bound.  Such
a mass matrix has long been regarded by phenomenologists as a
good starting point for understanding the fermion mass hierarchy 
and mixing \cite{Fritsch,Harari}.

This same vector $\balpha$ which appears in the mass matrices of 
quarks and leptons gets coupled to the strong framon $\Phi$ in 
the framon potential (\ref{VPhi}) simply by virtue of the double 
invariance required under both the local and global symmetries
via the so-called $\nu_2$ term.  Minimization of this potential
(\S3) gives a degenerate vacuum which depends on $\balpha$.  
Hence, if the vacuum changes with scale the vector $\balpha$ will 
change also (i.e., it rotates).

An explicit sample calculation carried out in \S4---6 shows that 
under renormalization in the strong sector, the vacuum changes 
with the renormalization scale $\mu$.  It then follows that 
$\balpha$ will rotate with $\mu$.  This rotation is a matter 
only of the vacuum, hence universal, i.e., independent of the 
fermion type in the mass matrix (\ref{mfact}) of which $\balpha$ 
appears.  The rotation is found (\S7) further to have fixed 
points at $\mu = \infty$ and $\mu = 0$.

The mass matrix (\ref{mfact}) thus possesses all the properties
(i.e., rank-one, rotating, universal and endowed with fixed 
points) which have been identified in an earlier analysis
\cite{r2m2} as needed to reproduce the hierarchical mass and
mixing patterns observed in experiment.  And, in common to all
such rotation schemes, it offers also a solution to the strong 
CP problem, transforming the theta-angle there into a CP-violating 
phase in the CKM matrix, giving a Jarlskog invariant of the 
appropriate order of magnitude for $\theta$ of order unity (\S9).  
These observations are 
not affected by the appearance (\S8) of a non-flat scale-dependent
metric in generation space in spite of its potential significance 
in theory.

It seems thus that simply by implementing the idea of framing, i.e. 
promoting frame vectors into dynamical variables, an idea borrowed 
from gravity, one seems to have already gone quite some way towards 
understanding the unusual features of the standard model which have 
so far been taken for granted.  One has yet to see whether the 
mass and mixing parameters observed in experiment can actually 
be accommodated in the FSM, and whether the model might lead to some 
consequences at variance with experiment elsewhere, the latter in relation to the strong framons in particular, as mentioned at the end of \S4.  Some 
work has been done already towards those ends, which we hope
to report later.  But to answer these questions with 
confidence will clearly be a long process which will ultimately 
require the participation and scrutiny of the community.

For conclusion, a word of comparison between the present FSM with 
other models or theories probing what underlies the standard model 
may be in order.  Compared with models or theories of the type 
known as beyond the standard model (BSM), the FSM is obviously much 
more modest both in scope and in aim.  For example, superstring 
theory, the prime example of the BSM theories, starts with higher 
dimensions both of space-time and of the fundamental object (i.e.,  
from point particles to strings or branes) and extensions of the 
gauge symmetry (e.g., to SUSY), while the FSM remains in 4-dimensional 
space-time with point particles and the same local gauge symmetry 
$su(3) \times su(2) \times u(1)$ as the standard model 
itself.  And while superstring theory opens up a new world with 
almost boundless implications way beyond the confines of particle 
physics, in cosmology and cosmogony etc., the aims of the FSM remain 
within particle physics, at least for the present.  The virtue 
of a limited scope, however, is economy, so that for example the FSM, 
by explaining the mass hierarchy of fermions and 
their mixing patterns, can look to reducing, and even calculating, 
some of the many empirical parameters of the standard model in 
the future; whereas in BSM theories the number of parameters 
tends further to increase (in SUSY alone, there are already more 
than a hundred).  Nevertheless, there is no obvious contradiction 
of the tenets of the FSM with any BSM theories, nor is there 
any obvious obstacle in incorporating the FSM into those larger 
theories, i.e., if ever one so desires.

However, perhaps the most distinguishing (some might even say 
revolutionary) feature of the framed standard model is its 
suggestion that the origin of all those baffling intricacies in 
flavour physics is to be found not in the far ultraviolet region 
as most theories would advocate but at the energy accessible 
already to us today, only still unrecognized by us because it is hidden 
cleverly by nature from our view.  Fermion generation itself is
said to be the dual of colour, while the rotation of the fermion 
mass matrix, which is thought to lead to both mixing and the mass 
hierarchy, is seen in \S3---\S6 to be driven by hadronic interactions. 
And even the CP-violating phase in the CKM matrix is assigned a 
hadronic origin \S9 in the theta-angle term of the old strong 
CP problem.  If that is indeed the case, then it can in future
lead to a phenomenological bonanza, for the tests on its tenets 
can no longer be deferred to infinity as they can be for some 
other theories, but will have to be confronted by us today.

\vspace{1cm}

We are greatly indebted to James Bjorken for many exchanges over
the last two years on the subject of mass matrix rotation.  
Although his approach to deriving rotation is quite different from the
FSM here, he has given us great 
encouragement on our approach and 
has sharpened considerably our own ideas by probing us with some 
questions that we should have asked ourselves but did not.   

\vspace{1cm}

\appendix

\begin{flushleft}
{\bf\large Appendix A}
\end{flushleft}

The Higgs mass spectrum can be found straightforwardly by computing
the second derivatives of the framon potential $V[\Phi]$.  It is not 
needed in this paper but will be useful in future applications.  The
resulting $10 \times 10$ matrix is block diagonal, with the lower $6
\times 6$ block actually diagonal.  With a little more manipulation (by
elementary row operations), we can further reduce the upper block, so
that in the end we get

\begin{equation}
M_{H}=
\pmatrix{
4\lambda_{W}\zeta_{W}^{2} & 2\zeta_{W}\zeta_{S}(\nu_{1}-\nu_{2})
\sqrt{\frac{1+2R}{3}} &
2\sqrt{2}\zeta_{W}\zeta_{S}\nu_{1}\sqrt{\frac{1-R}{3}} & 0 \cr
\ast & 4(\kappa_{S}+\lambda_{S})\zeta_{S}^{2}\left(\frac{1+2R}{3}\right) & 
4\sqrt{2}\lambda_{S}\zeta_{S}^{2}\frac{\sqrt{(1+2R)(1-R)}}{3} & 0 \cr
\ast & \ast &
4(\kappa_{S}+2\lambda_{S})\zeta_{S}^{2}\left(\frac{1-R}{3}
\right) & 0 \cr
0 & 0 & 0 & D
}
\label{higgsmassmatrix}
\end{equation}
where 
\begin{equation}
D=\kappa_{S}\zeta_{S}^2
\pmatrix{
4(\frac{1-R}{3}) & 0 & 0 & 0 & 0 & 0 & 0 \cr
0 &4(\frac{1-R}{3}) & 0 & 0 & 0 & 0 & 0 \cr
0 & 0 & 4(\frac{1-R}{3}) & 0 & 0 & 0 & 0 \cr
0 & 0 & 0 & 2(\frac{2+R}{3}) & 0 & 0 & 0 \cr
0 & 0 & 0 & 0 & 2(\frac{2+R}{3}) & 0 & 0 \cr
0 & 0 & 0 & 0 & 0 & 2(\frac{2+R}{3}) & 0 \cr
0 & 0 & 0 & 0 & 0 & 0 & 2(\frac{2+R}{3}) \cr
},
\end{equation}
where an $\ast$ denotes the corresponding symmetric entry, and where the first row (and column) corresponds to the electroweak state $h$.

The fact that if we wish to find the Higgs masses we need only
diagonalize a $3 \times 3$ matrix makes it theoretically possible.
However, short of actually finding the eigenvalues (other than 
numerically), which involves solving cubic equations, we can 
usefully find conditions for which the eigenvalues are positive.

An elementary result from linear algebra says that a real symmetric  
$3 \times 3$ matrix 
\begin{equation}
\pmatrix{
a_{11} & a_{12} & a_{13} \cr
\ast & a_{22} & a_{23} \cr
\ast & \ast & a_{33} \cr
}
\end{equation}
has positive eigenvalues if and only if 
\begin{enumerate}
\item $a_{11} >0$ ,
\item  $\det \pmatrix{
a_{11} & a_{12} \cr
\ast & a_{22}  \cr
} >0 $,
\item $\det  \pmatrix{
a_{11} & a_{12} & a_{13} \cr
\ast & a_{22} & a_{23} \cr
\ast & \ast & a_{33} \cr
} > 0$.
\end{enumerate}

Assuming all coupling constants (except possibly $\nu_1,\nu_2$)
to be positive, and also $ -1/2
< R <1$, we find necessary and sufficient conditions 
on the coupling constants for positive 
Higgs masses:
\begin{eqnarray}
4 \lambda_W (\lambda_S+\kappa_S)& > & (\nu_1 -\nu_2)^2, \nonumber \\
4 \lambda_W \kappa_S (\kappa_S +3 \lambda_S) & > & \kappa_S (\nu_1 -
\nu_2)^2 + 2 \lambda_S \nu_2^2 + 2 \kappa_S \nu_1^2.
\end{eqnarray}

These conditions are satisfied when either all coupling
constants are 1, or when $\nu_i$ are
small compared to the other coupling constants.

We can replace the above two necessary and sufficient conditions
  by two neater sufficient conditions 
\begin{eqnarray}
  4 \lambda_W (\lambda_S+\kappa_S)& >& (\nu_1 -\nu_2)^2,\nonumber \\
  4 \lambda_W \lambda_S \kappa_S & > & \kappa_S \nu_1^2 + \lambda_S
  \nu_2^2 .
\end{eqnarray}

Note that the above quoted result about positivity of eigenvalues is a direct
generalization of the conditions for a local minimum of a surface, in
dimension 2, and has a straightforward generalization to dimension $n>3$.  

\vspace{1cm}

\appendix

\begin{flushleft}
{\bf\large Appendix B}
\end{flushleft}

The space of degenerate vacua is parametrized by $\widetilde{SU}
(3)$, but at the moment we are interested in the classes of vacua
corresponding to various \balpha.  Also, so far we have essentially
only real \balpha, so that the above metric is worked out implicitly
in three real dimensions.

However, as \balpha\ runs, we should take into account the parameter
$t=\ln \mu^2$, so that we are really not in $\bbr^3$, but in $\bbr^3 \times
\bbr$.   This is clear if we think of where the
RGE curve lies.  In fact, we should think of a metric which is flat in
the $t$ or $0$ direction, and that the other components depend only on
this coordinate, so that the $t=$ constant surfaces have constant
metric.  This is a (Riemannian) 
metric of Bianchi Type I, the simplest type. 

Below we shall work out explicitly, by calculating the Christoffel
symbols, the parallel transport matrix (\ref{paratrans}). We shall
do so, for simplicity, only for the case $A = 1$, i.e., when there is no 
rotation with scale.  In the following we use superscripts to denote 
vector components, $\balpha=(\alpha^1,\alpha^2,\alpha^3)$.

Take coordinates $(t,x,y,z)$ indexed by $(0,1,2,3)$.  We start with 
the metric

\begin{equation}
\tilde{G}^{-1}=\left(\begin{array}{cccc}
1&0&0&0\\
0&\frac{1}{1+2R}&0&0\\
0&0&\frac{1}{1-R}&0\\
0&0&0&\frac{1}{1-R}\end{array}\right).
\label{4metric}
\end{equation}
From this we find, using equation (\ref{Christoffel}), the non-zero 
Christoffel symbols, for $i=2,3$,

\begin{eqnarray}
\Gamma^0_{11}&=&-\frac{1}{2}\frac{\partial}{\partial t}\left(
\frac{1}{1+2R}\right),\\
\Gamma^0_{ii}&=&-\frac{1}{2}\frac{\partial}{\partial t}\left(
\frac{1}{1-R}\right),\\
\Gamma^1_{10}=\Gamma^1_{01}&=&\frac{1}{2}(1+2R)\frac{\partial}{\partial
  t}\left(\frac{1}{1+2R}\right),\\
\Gamma^i_{i0}=\Gamma^i_{0i}&=&\frac{1}{2}(1-R)\frac{\partial}{\partial
  t}\left(\frac{1}{1-R}\right).
\end{eqnarray}
We have the covariant derivative of a vector $\balpha$ in the $t$ 
direction given by

\begin{equation}
\nabla_t \balpha = \left(\frac{\partial \alpha^i}{\partial t}+
\alpha^k\Gamma^i_{0k}\right)e_i.
\end{equation}
If we now consider just the $x$ component of this we find, using 
$\Gamma^i_{0j}=0$ for $i \neq j$,

\begin{equation}
(\nabla_t \balpha)^1 = \frac{\partial \alpha^1}{\partial t}+
\alpha^1\Gamma^1_{01}.
\end{equation}
Now parallel transport means that $(\nabla_t\balpha)^1=0$ so

\begin{equation}
\frac{\partial\alpha^1}{\partial t} = -\frac{1}{2}(1+2R)
\frac{\partial}{\partial t}\left(\frac{1}{1+2R}\right)\alpha^1.
\end{equation}
Since $\balpha$ and $R$ are functions of $t$ only the partial
derivatives 
are in fact total derivatives and we can now easily integrate
\begin{eqnarray}
\int^{t'}_{t}\frac{d}{dt}\left(\ln\alpha^1\right)dt&=&\int^{t'}_{t}
\frac{d}{dt}(\ln(1+2R)^{\frac{1}{2}})dt,\\
\left(\frac{{\alpha^1}'}{\alpha^1}\right)&=&\left(\frac{1+2R'}{1+2R}
\right)^{\frac{1}{2}}.
\end{eqnarray}
Similarly if we consider the $y$ and $z$ components of the covariant
derivative we find

\begin{eqnarray}
\frac{\partial\alpha^2}{\partial t} &=& -\frac{1}{2}(1-R)
\frac{\partial}{\partial t}\left(\frac{1}{1-R}\right)\alpha^2,\\
\frac{\partial\alpha^3}{\partial t} &=& -\frac{1}{2}(1-R)
\frac{\partial}{\partial t}\left(\frac{1}{1-R}\right)\alpha^3,
\end{eqnarray}
which we can integrate to give

\begin{equation}
\left(\frac{{\alpha^2}'}{\alpha^2}\right)=\left(
\frac{{\alpha^3}'}{\alpha^3}\right)=\left(\frac{1-R'}{1-R}
\right)^{\frac{1}{2}}.
\end{equation}
We can now write parallel transport (for $A=1$) in the $t$ direction as

\begin{equation}
\balpha'=\left(
\begin{array}{ccc}\frac{P'}{P}&0&0\\0&\frac{Q'}{Q}&0\\0&0&
\frac{Q'}{Q}\end{array}\right)\balpha.
\end{equation}

\end{document}